\documentclass[journal,onecolumn,11pt]{IEEEtran}
\pdfoutput=1
\usepackage{amssymb}
\usepackage{amsmath}
\usepackage{graphicx}
\usepackage{xcolor}
\usepackage{xspace}
\usepackage{enumerate}
\usepackage{multicol}

\usepackage{amsfonts}
\usepackage{url} 
\usepackage[colorlinks=true]{hyperref} 
\usepackage{balance}
\usepackage[lined,ruled,linesnumbered]{algorithm2e}
\usepackage{subfig}
\usepackage{amsthm}
\theoremstyle{definition}

\def\figspec{}

\definecolor{NewTextBG}{rgb}{0.8,0.8,1.00}

\newcommand{\iter}[1]{^{(#1)}}
\newcommand{\BlackBox}{\rule{1.5ex}{1.5ex}}  

\newcommand{\opt}[1]{\ensuremath{\hat{#1}}}

\def\reals{\ensuremath{\mathbb{R}}}

\renewcommand{\vec}[1]{\ensuremath{\mathbf{\MakeLowercase{#1}}}}
\newcommand{\mat}[1]{\ensuremath{\mathbf{\MakeUppercase{#1}}}}
\newcommand{\rv}[1]{\ensuremath{\MakeUppercase{#1}}}
\newcommand{\dpdf}[1]{\mathrm{\ensuremath{\MakeUppercase{#1}}}}
\newcommand{\cpdf}[1]{\mathrm{\ensuremath{\MakeLowercase{#1}}}}
\newcommand{\st}{\ensuremath{\quad\mathrm{s.t.}\quad}}
\newcommand{\norm}[1]{\ensuremath{\left\|#1\right\|}}
\newcommand{\quant}[1]{\ensuremath{\left[#1\right]}}
\newcommand{\support}[1]{\mathrm{supp}(#1)}

\def\rank{\mathrm{rank}}
\newcommand{\fun}[1]{\mathrm{#1}}
\newcommand{\cost}[1]{\ensuremath{\ell_{#1}}\xspace}
\newcommand{\abs}[1]{\ensuremath{\left|#1\right|}}
\newcommand{\setdef}[1]{\ensuremath{\left\{#1\right\}}}

\newcommand{\svec}[1]{_{[#1]}}
\def\transp{^\intercal}
\newcommand{\refeq}[1]{(\ref{#1})}

\def\Gaussian{\ensuremath{\mathcal{N}}}
\newcommand{\GaussianPDF}[2][\sigma]{\ensuremath{\frac{1}{\sqrt{2\pi#1^2}}e^{-\frac{#2^1}{2#1^2}} }}
\def\Exponential{\ensuremath{\mathrm{Exp}}}
\def\Bernoulli{\ensuremath{\mathrm{Ber}}}
\def\Laplacian{\ensuremath{\mathrm{Lap}}}
\newcommand{\LaplacianPDF}[2][\theta]{\ensuremath{\frac{1}{2#1}e^{-\frac{|#2|}{#1}} }}
\def\Indicator{\ensuremath{\mathbf{1}}}

\def\sgn{\mathrm{sgn}}

\def\defeq{:=}
\def\assign{\leftarrow}

\def\inv{^{-1}}

\def\model{M}
\def\dict{d}
\def\dictm{\mat{\dict}}
\def\dictv{\vec{\dict}}

\def\dpred{b}

\def\dpredv{\vec{\dpred}}

\def\data{y}
\def\datam{\mat{\data}}
\def\datav{\vec{\data}}

\def\err{e}
\def\errm{\mat{\err}}
\def\errv{\vec{\err}}
\def\errrv{\rv{\err}}

\def\noiserv{N}

\def\coef{a}
\def\coefm{\mat{\coef}}
\def\coefv{\vec{\coef}}
\def\coefrv{\rv{\coef}}

\def\supp{z}
\def\suppm{\mat{\supp}}
\def\suppv{\vec{\supp}}
\def\supprv{\rv{\supp}}

\def\sign{s}

\def\signv{\vec{\sign}}
\def\signrv{\rv{\sign}}

\def\corr{g}

\def\corrv{\vec{\corr}}

\def\canon{\omega}

\def\val{v}

\def\valv{\vec{\val}}
\def\valrv{\rv{\val}}

\def\aux{u}

\def\auxv{\vec{\aux}}

\def\ndims{m} 
\def\natoms{p}
\def\nsamples{n}
\def\nclasses{c}

\def\datavspace{\reals^{\ndims}}
\def\coefvspace{\reals^{\natoms}}
\def\suppvspace{\setdef{0,1}^{\natoms}}

\def\dictmspace{\reals^{\ndims{\times}\natoms}}

\def\di{i} 
\def\si{j} 
\def\ai{k} 

\def\supsize{\norm{\coefv}_0}
\def\maxnz{\gamma}

\def\modelclass{\mathcal{M}}
\title{An MDL framework for sparse coding and dictionary learning}
\author{Ignacio Ram\'{i}rez and Guillermo Sapiro\thanks{Department of Electrical and Computer Engineering, University of Minnesota. Work supported by NSF, NGA, ARO, ONR, DARPA and NSSEFF.}}
\begin{document}
%
\maketitle
\begin{abstract}
  The power of sparse signal modeling with learned over-complete
  dictionaries has been demonstrated in a variety of applications and
  fields, from signal processing to statistical inference and machine
  learning. However, the statistical properties of these models, such as
  under-fitting or over-fitting \emph{given} sets of data, are still not
  well characterized in the literature. As a result, the success of sparse
  modeling depends on hand-tuning critical parameters for each data and
  application. This work aims at addressing this by providing a practical
  and objective characterization of sparse models by means of the Minimum
  Description Length (MDL) principle -- a well established
  information-theoretic approach to model selection in statistical
  inference. The resulting framework derives a family of efficient sparse
  coding and dictionary learning algorithms which, by virtue of the MDL
  principle, are completely parameter free. Furthermore, such framework
  allows to incorporate additional prior information to existing models,
  such as Markovian dependencies, or to define completely new problem
  formulations, including in the matrix analysis area, in a natural way.
  These virtues will be demonstrated with parameter-free algorithms for the
  classic image denoising and classification problems, and for low-rank matrix
  recovery in video applications.
\end{abstract}

\section{Introduction}
\label{sec:introduction}

A \emph{sparse model} is one in which signals of a given type $\datav
\in \datavspace$ can be represented accurately as sparse linear combinations
of the columns (atoms) of a learned dictionary $\dictm \in \dictmspace$,
$\datav = \dictm\coefv + \errv,$ where by accurate we mean that
$\norm{\errv}\ll \norm{\datav}$ (in some norm), and by sparse we mean that
the number of non-zero elements in $\coefv$, denoted by $\norm{\coefv}_0$,
is small compared to its dimension $\natoms$. These concepts will be
formalized in the next section.

Such models, especially when $\dictm$ is learned from training samples, are
by now a well established tool in a variety of fields and applications,
see~\cite{bruckstein09,rubinstein10ieee,wright10ieee} for recent reviews.

When sparsity is a modeling device and not an hypothesis about the nature of
the analyzed signals, parameters such as the \emph{desired} sparsity in the
solutions, or the size $\natoms$ of the dictionaries to be learned, play a
critical role in the effectiveness of sparse models for the data and tasks
at hand. However, lacking theoretical guidelines for such parameters,
published applications based on learned sparse models often rely on either
cross-validation or ad-hoc methods for determining such critical parameters
(an exception for example being the Bayesian approach,
e.g.,~\cite{carin11}).  Clearly, such techniques can be impractical and/or
ineffective in many cases. This in turn hinders the further application of
such models to new types of data and applications, or their evolution into
different, possibly more sophisticated, models.

At the bottom of the aforementioned problem lie fundamental questions such
as: \emph{How rich or complex is a sparse model? How does this depend on the
required sparsity of the solutions, or the size of the dictionaries? What is
the best model for a given data class and a given task?}

The general problem of answering such questions and, in particular, the
latter, is known as \emph{model selection}. Popular model selection
techniques such as Akaike's Information Criterion
(AIC)~\cite{akaike74}, Bayes Information Criterion
(BIC)~\cite{schwartz78}, and the Minimum Description Length principle
(MDL)~\cite{rissanen78,rissanen84,barron98}, work by building a cost
function which balances a measure of \emph{goodness of fit} with one of \emph{model
complexity}, and search for a model that minimizes such cost. In this sense,
these tools can be regarded as practical implementations of the Occam's
razor principle, which states that, given two (equally accurate)
descriptions for a given phenomenon, the simpler one is usually the best.

In the Minimum Description Length principle, given a family or model class
$\modelclass$ of candidate models indexed by a parameter $\model$, and a data
sample $\datav$, the best model $\opt{\model} \in \modelclass$ is the one
that can be used to describe $\datav$ completely (including the parameters
$\model$ themselves) with the fewest number of bits,
\begin{equation}
 \opt{\model} = \arg\min_{\model \in \modelclass} L(\datav,\model),
\label{eq:mdl-gen}
\end{equation}
where $L(\datav,\model)$ is a \emph{codelength assignment function} which
defines the theoretical codelength required to describe $(\datav,\model)$
\emph{uniquely}, and which is a key component of any MDL-based
framework. The underlying idea of MDL is that \emph{compressibility is a
  good indirect way of measuring the ability of a model to capture
  regularity from the data}. Common practice in MDL uses the \emph{Ideal
  Shannon Codelength Assignment} \cite[Chapter~5]{cover06} to define
$L(\datav,\model)$ in terms of a \emph{probability assignment}
$\dpdf{p}(\datav,\model)$ as $L(\datav,\model)=-\log
\dpdf{p}(\datav,\model)$ (all logarithms will be assumed on base $2$
hereafter). In this way, the problem of choosing $L(\cdot)$ becomes one of
choosing a suitable probability model for $(\datav,\model)$. Note here how
MDL considers probability models not as a statement about the true nature of
the data, but only as a modeling tool.  If we now write
$\dpdf{p}(\datav,\model)=\dpdf{p}(\datav|\model)\dpdf{p}(\model)$, we obtain
the more familiar \emph{penalized likelihood} form,
\begin{equation}
 \opt{\model} = \arg\min_{\model \in \modelclass} -\log \dpdf{P}(\datav|\model) -\log \dpdf{P}(\model),
\label{eq:mdl-twoparts}
\end{equation}
with $-\log \dpdf{P}(\model)$ representing the model complexity, or
\emph{model cost}, term. 

The use of MDL for sparse signal modeling has been explored for example in
the context of wavelet-based denoising (where $\model=\coefv \in
\reals^\ndims$, $\natoms=\ndims$ and $\dictm \in
\reals^{\ndims{\times}\ndims}$ is \emph{fixed}) of images corrupted by
additive white Gaussian noise (AWGN) \cite{saito94, krim95, moulin99,
  rissanen00, roos09}.  In \cite{saito94,krim95,moulin99}, the data is
described using \refeq{eq:mdl-twoparts} with $-\log\dpdf{p}(\datav|\coefv)$
assumed to be \emph{solely due to noise}, and an $L(\coefv)$ term which
exploits sparsity,
\begin{equation}
  \opt{\coefv} = \arg\min_{\coefv \in \modelclass} \frac{1}{2\sigma^2_e}\norm{\datav-\dictm\coefv}_2^2 + L(\coefv).
\end{equation}
Here the first term corresponds to the ideal codelength, up to a constant,
of an IID Gaussian sequence of zero mean and known variance
$\sigma^2_e$. The difference between \cite{saito94, krim95, moulin99} lies
in the definition of $L(\coefv)$.  The line of work \cite{rissanen00,roos09}
follows the modern MDL approach by using sophisticated tools from coding
theory, the so called \emph{one-part universal codes}, which encodes
$(\datav,\coefv)$ jointly, and reduces the arbitrariness in defining
$L(\coefv)$. However, such tools can only be applied for certain choices of
$\dpdf{p}(\datav|\coefv)$ and $\dpdf{p}(\coefv)$. In the case of
\cite{rissanen00,roos09}, the choice is to use continuous Gaussian models
for both. As Gaussian models are \emph{not well suited to the typically
  observed statistical properties of such data}, the performance of the
resulting denoisers for example is very poor compared to the current
state-of-the-art.

The present work extends and/or improves on the aforementioned work in the
following ways:\footnote{This paper extends preliminary results reported in
  \cite{ramirez11icassp}. In particular, new dictionary learning algorithms
  are developed which include $\cost{1}$ atom regularization, forward and
  backward dictionary size adaptation. We also develop a new model for the
  low-rank matrix approximation problem.}
\begin{itemize}
\item MDL-based sparse coding is extended to the case of
  \emph{non-orthonormal, possibly over-complete and learned dictionaries}
  $\dictm$. As we will see in Section~\ref{sec:encoding-algorithms}, this
  extension, critical to deal with modern, very successful sparse modeling
  approaches, poses not only new design problems but also significant
  computational challenges compared to the orthonormal case.
\item Efficient codelengths (probability distributions) for the different
  components to encode (error, coefficients, dictionary) are obtained by
  \emph{applying universal coding schemes to priors that are suited to the
    typically observed statistics of such data.}
\item As a particular point of the above item, systematic model-fit
  deviations are naturally taken into account in
  $\dpdf{p}(\datav|\coefv)$. The resulting fitting terms fall into the
  category of robust estimators (see~\cite{huber64}), thus marrying robust
  statistics with information theory and with sparse modeling (dictionary
  learning).
\item We comply with the basic MDL sanity check, meaning, that \emph{the
  theoretical codelengths obtained are smaller than a ``raw''
  description of the data}. We do so by including quantization in our models,
  and treating its effect rigorously.
\item Dictionary learning within the MDL framework allows us to
  \emph{optimize both the number of atoms $\natoms$, as well as their
    structure}, resulting in a natural and objective form of regularization
  for $\dictm$.
\item Structure is naturally added to the sparse models in the form of
  Markovian dependencies between adjacent data samples. We also show an
  extension of the model to the problem of low-rank matrix completion.
\end{itemize}

As a result of the above features, we obtain for the first time an
MDL-based, parameter-free framework for signal modeling that is able to
yield state-of-the-art results.

At the theoretical level, this brings us a step closer to the fundamental
understanding of \emph{learned} sparse models and brings a different
perspective, that of MDL, into the sparse modeling world.

The remainder of this paper is organized as follows. Sparse models, and the
associated notation, are described in detail in
Section~\ref{sec:background}. Section~\ref{sec:mdl-model-selection}
introduces MDL, and its application to sparse models. In
Section~\ref{sec:encoding-scheme} we present the probability models used to
assign codelengths to different parts of the encoded data, while
sections~\ref{sec:encoding-algorithms} and~\ref{sec:learning-algorithms}
describe the actual sparse coding and dictionary learning algorithms
developed. Experimental results follow in Section~\ref{sec:results}, and the
paper is concluded in Section~\ref{sec:conclusion}.

\section{Background on sparse modeling}
\label{sec:background}

Assume we are given $\nsamples$ $\ndims$-dimensional data samples ordered as
columns of a matrix $\datam=[\datav_1|\datav_2|\ldots|\datav_\nsamples] \in
\reals^{\ndims{\times}\nsamples}$. Consider a linear model for $\datam$,
$\datam = \dictm\coefm + \errm,$ where
$\dictm=[\dictv_1|\dictv_2|\ldots|\dictv_\natoms]$ is an
$\ndims{\times}\natoms$ dictionary consisting of $\natoms$ atoms,
$\coefm=[\coefv_1|\coefv_2|\ldots|\coefv_\nsamples] \in
\reals^{\natoms{\times}\nsamples}$ is a matrix of coefficients where each
$\si$-th column $\coefv_\si$ specifies the linear combination of columns of
$\dictm$ that approximates $\datav_\si$, and
$\errm=[\errv_1|\errv_2|\ldots|\errv_\nsamples] \in
\reals^{\ndims{\times}\nsamples}$ is a matrix of approximation errors.  

We define the support, or active set, of a vector $\coefv \in \coefvspace$
as $\support{\coefv}= \setdef{\ai:\coefv_\ai \neq 0}$. Let
$\Gamma=\support{\coefv}$. We also represent the support of $\coefv$ as a
binary vector $\suppv \in \suppvspace$ such that $\supp_i=1$ for $i \in
\Gamma$, and $0$ otherwise. We refer to the sub-vector in
$\reals^{|\Gamma|}$ of non-zero elements of $\coefv$ as either
$\coefv\svec{\Gamma}$ or $\coefv\svec{\suppv}$. Both conventions are
extended to refer to sets of columns of matrices, for example,
$\dictm\svec{\Gamma}$ is the matrix formed by the $|\Gamma|$ columns of
$\dictm$ indexed by $\Gamma$. We will use the pseudo-norm
$\norm{\coefv}_0\defeq |\Gamma| =\sum\suppv$ to denote the number of
non-zero elements of $\coefv$.  We say that the model is \emph{sparse} if we
can achieve $\norm{\errv_\si}_2 \ll \norm{\datav_\si}_2$ and
$\norm{\coefv}_0 \ll \natoms$ simultaneously for all or most
$\si=1,\ldots,\nsamples$.

The result of quantizing a real-valued variable $\data$ to precision
$\delta$ is denoted by $\quant{\data}_\delta$. This notation is extended to
denote element-wise quantization of vector (e.g., $\quant{\errv}$) and
matrix operands (e.g., $\quant{\errm}$).

\subsection{Sparse coding}
\label{sec:background:coding}

One possible form of expressing the \emph{sparse coding problem} is given
by
\begin{equation}
{\opt{\coefv}_\si} \!=\! \arg\min_{\auxv \in \coefvspace} 
\norm{\datav_\si\!-\!\dictm\auxv}_2
\st\norm{\auxv}_0  \leq \maxnz,
\label{eq:l0-sparse-coding}%
\end{equation}
where $\maxnz \ll \natoms$ indicates the desired \emph{sparsity level} of
the solution. Since problem \refeq{eq:l0-sparse-coding} is non-convex and
NP-hard, approximate solutions are sought. This is done either by using
greedy methods such as Matching Pursuit (MP) \cite{mallat93}, or by solving a
convex approximation to \refeq{eq:l0-sparse-coding}, commonly known as the
\emph{lasso}~\cite{tibshirani96},
\begin{equation}
\opt{\coefv}_\si = \arg\min_{\auxv \in \coefvspace} \frac{1}{2}\norm{\datav_\si-\dictm\auxv}_2 
\!\!\st\! \norm{\auxv}_1 \leq \tau.
\label{eq:l1-sparse-coding-lasso}
\end{equation}
There exists a body of results showing that, under certain conditions on
$\maxnz$ and $\dictm$, the problem \refeq{eq:l0-sparse-coding} can be solved
exactly via \refeq{eq:l1-sparse-coding-lasso} or MP (see for example~\cite{bruckstein09,candes06}).
In other cases, the objective is not to solve \refeq{eq:l0-sparse-coding},
but to guarantee some property of the estimated $\opt{\coefv}_\si$. For
example, in the above mentioned case of AWGN denoising in the wavelets
domain, the parameter $\tau$ can be chosen so that the resulting estimators
are universally optimal with respect to some class of signals
\cite{donoho94}. However, if $\dictm$ is arbitrary, no such choice
exists. Also, if $\dictm$ is orthonormal, the problem
\refeq{eq:l1-sparse-coding-lasso} admits a closed form solution
obtained via the so-called \emph{soft thresholding}
\cite{donoho94}. However, again, for general $\dictm$, no such solution
exists, and the search for efficient algorithms has been a hot topic
recently, e.g., \cite{friedman08,beck09siam,efron04}.

\subsection{Dictionary learning}
\label{sec:background:learning}
When $\dictm$ is an optimization variable, we refer to the resulting problem
as \emph{dictionary learning}:
\begin{equation}
  (\opt\coefm,\opt\dictm) = 
\arg\min_{\coefm,\dictm} \sum_{\si=1}^{\nsamples}
{\frac{1}{2}\norm{\datav_\si-\dictm\coefv_\si}_2^2\;
\st \norm{\coefv_\si}_r}\leq\tau\,\forall\,j,\;\;
 \norm{\dictv_k}_2 \leq 1\,\forall k,
\label{eq:traditional-dictionary-learning}
\end{equation}
with $0 \leq r \leq 1$. The constraint $\norm{\dictv_k}_2 \leq
1\,,\;k=1,\ldots,\natoms$, is necessary to avoid an arbitrary decrease of the
cost function by setting $\dictm \assign \alpha\dictm$, $\coefm \assign
\frac{1}{\alpha}\coefm$, for any $\alpha > 1$. 
The cost function in \refeq{eq:traditional-dictionary-learning} is
non-convex in $(\coefm,\dictm)$, so that only local convergence can be
guaranteed. This is usually achieved using alternate optimization in
$\dictm$ and $\coefm$. See for example \cite{aharon06,mairal10jmlr} and
references therein.

\subsection{Issues with traditional sparse models: a motivating example}
\label{sec:background:issues}

Consider the K-SVD-based~\cite{aharon06} sparse image restoration framework
\cite{mairal08a}. This is an \cost{0}-based dictionary learning framework,
which approximates \refeq{eq:traditional-dictionary-learning} for the case
$r=0$ by alternate minimization. In the case of image denoising, the general
procedure can be summarized as follows:

\begin{enumerate}
\item An initial, \emph{global} dictionary $\dictm_0$ is learned
  using training samples for the class of data to be processed (in this case
  small patches of natural images). The user must supply a patch width $w$,
  a dictionary size $\natoms$ and a value for $\tau$.
\item The noisy image is decomposed into overlapping $w{\times}w$ patches (one patch
  per pixel of the image), and its noisy patches are used to further adapt $\dictm$ using the following
  \emph{denoising} variant of \refeq{eq:traditional-dictionary-learning},
\begin{align}
  (\opt\dictm,\opt\coefm) = 
\arg\min_{\dictm,\coefm} \sum_{\si=1}^{\nsamples}
   \norm{\coefv_\si}_0
\,,
\st 
&\frac{1}{2}\norm{\datav_\si-\dictm\coefv_\si}_2^2 \leq C\sigma^2\,,\;
&\norm{\dictv_k}_2 = 1\,,\;k=1,\ldots,\natoms.
\label{eq:traditional-dictionary-learning-denoising}
\end{align}
Here the user must further supply a constant $C$ (in \cite{mairal08a}, it is
$1.32$), the noise variance $\sigma^2$, and the number of iterations $J$ of
the optimization algorithm, which is usually kept
small to avoid over-fitting (the algorithm is \emph{not} allowed to converge).
\item The final image is constructed by assembling the patches in
  $\opt\datam=\opt\dictm\opt\coefm$ into the corresponding original
  positions of the image. The final pixel value at each location is an
  average of all the patches to which it belongs, plus a small fraction $0
  \leq \lambda \leq 1$ of the original noisy pixels ($\lambda=30/\sigma$ in \cite{mairal08a}).

\end{enumerate}

Despite the good results obtained for natural
images, several aspects of this method are not satisfactory:
\begin{itemize}
\item Several parameters ($w$, $\natoms$, $\tau$, $C$, $J$, $\lambda$) need to be tuned. \emph{There
  is no interpretation, and therefore no justifiable choice for these
  parameters, other than maximizing the empirical performance of the
  algorithm (according to some metric, in this case PSNR) for the data at
  hand.}
\item The effect of such parameters on the result is shadowed by the effects
  of later stages of the algorithm and their associated parameters
  (e.g. overlapping patch averaging). \emph{There is no fundamental way to
    optimize each stage separately.}
\end{itemize}

As a partial remedy to the first problem, Bayesian sparse models were
developed (e.g., \cite{carin11}) where these parameters are assigned prior
distributions which are then learned from the data. However, this approach
still does not provide objective means to compare different models (with
different priors, for example). Further, the Bayesian technique implies
having to repeatedly solve possibly costly optimization problems, increasing
the computational burden of the application.

As mentioned in the introduction, this work proposes to address the above
practical issues, as well as to provide a new angle into dictionary
learning, by means of the MDL principle for model selection. The details on
how this is done are the subject of the following sections.

\section{Sparse model selection and MDL}
\label{sec:mdl-model-selection}

Given data $\datam$, a maximum support size $\maxnz$ and a dictionary size
$\natoms$, traditional sparse modeling provides means to estimate the best
model $\model=(\coefm,\dictm)$ for $\datam$ within the set
$\modelclass(\maxnz,\natoms)$ defined as
\begin{equation}
\modelclass(\maxnz,\natoms) \defeq \setdef{(\coefm,\dictm): \norm{\coefv_\si}_0 \leq \maxnz, \si=1,\ldots,\nsamples, \dictm \in \dictmspace }.
\label{eq:traditional-sparse-modeling}
\end{equation}
We call such set a \emph{sparse model class} with hyper-parameters
$(\gamma,\natoms)$. Such classes are nested in the following way: first, for
a fixed dictionary size $\natoms$ we have $\modelclass(\maxnz-1,\natoms)
\subset \modelclass(\maxnz,\natoms)$. Also, for fixed $\maxnz$, if we
consider $\modelclass(\maxnz,\natoms-1)$ to be a particular case of
$\modelclass(\maxnz,\natoms)$ where the $\natoms$-th atom is all-zeroes and
$\coef_{\natoms\si}=0,\,\forall j$, then we also have that
$\modelclass(\maxnz,\natoms-1) \subset \modelclass(\maxnz,\natoms)$. 

If one wants to choose the best model among all possible classes
$\modelclass(\maxnz,\natoms)$, the problem becomes one of \emph{model
  selection}.  The general objective of model selection tools is to define
an objective criterion for choosing such model. In particular, MDL model
selection uses codelength as such criterion. More specifically,
this means first computing the best model within each family as
\[
(\coefm(\maxnz,\natoms),\dictm(\maxnz,\natoms)) = \arg\min \{ L(\datam,\coefm,\dictm): {(\coefm,\dictm) \in \modelclass(\maxnz,\natoms)}\},
\]
and then choosing $(\opt{\maxnz},\opt{\natoms}) = \arg\min
\setdef{L(\datam,\coefm(\maxnz,\natoms),\dictm(\maxnz,\natoms)):
  0\leq\maxnz\leq\natoms,\,\natoms > 0}$.

When $\dictm$ is fixed, which is the case of sparse coding, the only model
parameter is $\coefm$, and we have $\natoms+1$ possible classes,
$\modelclass(\maxnz) = \setdef{\coefm: \norm{\coefv_\si}_0 \leq \maxnz,
  \si=1,\ldots,\nsamples}$, one for each $0 \leq \maxnz \leq \natoms$. If
each data sample $\datav_\si$ from $\datam$ is encoded independently, then,
as with traditional sparse coding (the framework can also be extended to
\emph{collaborative} models), the model selection problem can be broken into
$\nsamples$ sub-problems, one per sample, by redefining the model class
accordingly as $\modelclass(\maxnz) = \setdef{\coefv: \norm{\coefv}_0 \leq
  \maxnz}$. Clearly, in the latter case, the optimum $\maxnz$ can vary from
sample to sample.

Compared to the algorithm in Section~\ref{sec:background:issues}, we have a
\emph{fundamental, intrinsic} measure of the quality of each model, the
codelength $L(\datam,\coefm,\dictm)$, to guide our search through the
models, and which is unobscured from the effect of possible later stages of
the application. In contrast, there is no obvious intrinsic measure of
quality for models learned through \refeq{eq:traditional-sparse-modeling},
making comparisons between models learned for different parameters (patch
width $w$, regularization parameter $\tau$, norm $r$, constants $C,\lambda$) possible only in terms
of the observed results of the applications where they are embedded.
The second advantage of this framework  is that it allows  to select, in a
fundamental fashion, the best model parameters \emph{automatically}, thus
resulting in parameter-free algorithms.\footnote{For the case of image
  processing, the patch width $w$ is also a relevant parameter that could be
  automatically learned with the same MDL-based framework presented
  here. However, since it is specific to image processing, and due to space
  constraints and for clarity of the exposition, it will not be considered
  as part of the model selection problem hereafter.}

Such advantages will be of practical use only if the resulting computational
algorithms are not orders of magnitude slower than the traditional ones, and
efficient algorithms are a critical component of this framework, see
Section~\ref{sec:encoding-algorithms}.

\subsection{A brief introduction to MDL}
\label{sec:mdl-model-selection:intro}

For clarity of the presentation, in this section we will consider $\dictm$
fixed, and a single data sample $\datav$ to be encoded. The Minimum
Description Length principle was pioneered by Rissanen~\cite{rissanen78} in
what is called ``early MDL,'' and later refined by himself~\cite{rissanen84}
and other authors to form what is today known as ``modern MDL'' (see
\cite{grunwald07} for an up-to-date extensive reference on the subject). The
goal of MDL is to provide an objective criterion to select the model $M$,
out of a family of competing models $\mathcal{M}$, that gives the best
description of the \emph{given} data $\datav$. In this case of sparse coding
with fixed dictionary we have $M=\coefv$.

The main idea of MDL is that, the best model for the data at hand is the one
that is able to capture more \emph{regularity} from it. The more regularity
a model captures, the more succinct the description of the data will be
under that model (by avoiding redundancy in the description). Therefore, MDL
will select the best model as the one that produces the shortest (most
efficient) description of the data, which in our case is given by
$L(\datav,\coefv)$.

As mentioned in Section~\ref{sec:introduction}, MDL translates the problem
of choosing a codelength function $L(\cdot)$ to one of choosing probability
models by means of the ideal Shannon codelength assignment
$L(\datav,\coefv) = -\log \dpdf{P}(\datav,\coefv)$. It is common to extend
such ideal codelength to continuous random variables $x$ with probability
density function $\cpdf{p}(x)$ as $L(x) = -\log \cpdf{p}(x)$, by assuming
that they will be quantized with sufficient precision so that
\begin{equation}
\dpdf{P}(\quant{x}_\delta) \approx \cpdf{p}(x)\delta, 
\label{eq:approx-codelength}
\end{equation}
and disregarding the
constant term $-\log \delta$ in $L(x)$, as it is inconsequential for model
selection. However, in our framework, the optimum quantization levels will
often be large enough so that such approximations are no longer valid.

To produce a complete description of the data $\datav$, the best model
parameters $\opt{M}$ used to encode $\datav$ need to be included in the
description as well. If the only thing we know is that $\opt{M}$ belongs to
a given class $\modelclass$, then the cost of this description will depend
on how large and complex $\modelclass$ is.  MDL will penalize more those
models that come from larger (more complex) classes.  This is summarized in
one of the fundamental results underlying MDL~\cite{rissanen84}, which
establishes that the minimum number of bits required for encoding \emph{any}
data vector $\datav$ using a model from a class $\modelclass$ has the form
$
L_\modelclass(\datav) = \mathcal{L}_\modelclass(\datav) + \mathcal{C}(\modelclass),
$
where $\mathcal{L}_\modelclass(\datav)$ is called the \emph{stochastic
  complexity}, which depends only on the particular instance of $\datav$
being encoded, and $\mathcal{C}(\modelclass)$ is an unavoidable
\emph{parametric complexity} term, which depends \emph{solely} on the
structure, geometry, etc., of the model class $\modelclass$. 

In the initial version of MDL~\cite{rissanen78}, the parameter $\opt{M}$ was
first encoded separately using $L(\opt{M})$ bits, and then $\datav$ was
described given $\opt{M}$ using $L(\datav|\opt{M})$ bits, so that the
complete description of $\datav$ required $L(\datav|\opt{M}) + L(\opt{M})$
bits. This is called a \emph{two-parts code}. An asymptotic expression of
this MDL was developed in~\cite{rissanen78} which is equivalent to the BIC criterion~\cite{schwartz78},
only in the asymptotic regime.
%
%
As we will see next, modern MDL departs significantly from this two-parts
coding scheme.

\subsection{Modern MDL and universal coding}
\label{sec:mdl-model-selection:intro2}

The main difference between ``early''~\cite{rissanen78} and
``modern''~\cite{rissanen84,barron98} MDL is the introduction of
\emph{universal codes} as the main building blocks for computing
codelengths.  In a nutshell, universal coding can be regarded as an extension
of the original Shannon theory to the case where the probability model
$\dpdf{P}(\cdot)$ of the data to be encoded is not fully specified, but only
known to belong to a certain class of candidate probability models
$\modelclass$ (recall that classic Shannon theory assumes that
$\dpdf{P}(\cdot)$ is perfectly known).  For example, $\modelclass$ can be a
family of parametric distributions indexed by some parameter $M$. Akin to
Shannon theory, for an encoding scheme to be called \emph{universal}, the
codelengths it produces need to be optimal, in some sense, with respect to
the codelengths produced by all the models in $\modelclass$.

Various universality criteria exist. For example, consider the
\emph{codelength redundancy} of a model $\dpdf{Q}(\cdot)$,
$
\mathcal{R}(\datav;Q) = -\log \dpdf{Q}(\datav) - \left[\arg\min_{\dpdf{P} \in \modelclass} -\log \dpdf{P}(\datav)\right].
$
In words, this is the codelength overhead obtained with $\dpdf{Q}(\cdot)$
for describing an instance $\datav$, compared to the best model in
$\modelclass$ that could be picked for $\datav$, \emph{with hindsight} of
$\datav$. For example, if $\modelclass$ is a parametric family, such model
is given by the maximum likelihood (ML) estimator of $M$.
A model $\dpdf{Q}(\cdot)$ is called \emph{minimax universal}, if it
minimizes the \emph{worst case redundancy},
$
\mathcal{R}(Q) = \arg\max_{\datav \in \datavspace} \mathcal{R}(\datav;Q).
$
One of the main techniques in universal coding is \emph{one-part coding},
where the data $\datav$ and the best class parameter $\hat{M}$ are encoded
jointly.  Such codes are used in the line of work of ``MDL denoising'' due
to Rissanen and his collaborators~\cite{rissanen00,roos09}. However,
applying one-part codes at this level restricts the probability models to be
used.\footnote{In particular, those used in~\cite{rissanen00,roos09} are
  based on the Normalized Maximum Likelihood (NML) universal
  model~\cite{shtarkov87}, which requires closed-form MLE estimators for its
  evaluation, something that cannot be obtained for example with a Laplacian
  prior on $\coefv$ and non-orthogonal dictionaries.}  As a consequence, the
results obtained with this approach in such works are not competitive with
the state-of-the-art. Therefore, in this work, we maintain a two-parts
encoding scheme (or three parts, if $\dictm$ is to be encoded as well),
where we separately describe $\coefv$, $\dictm$, and $\datav$ given
$(\coefv,\dictm)$. We will however use universal codes to describe each of
these parts as efficiently as possible. Details on this are given in the
next section.

\section{Encoding scheme}
\label{sec:encoding-scheme}

We now define
the models and encoding schemes used to describe each of the parts that
comprise a sparse model for a data sample $\datav$; that is, the dictionary
$\dictm$, the coefficients $\coefv$, and the approximation error
$\errv=\datav-\dictm\coefv$ (which can include both the noise and the model
deviation), which can be regarded as the conditional description of $\datav$
given the model parameters $(\coefv,\dictm)$.  The result will be a cost
function $L(\datav)$ of the form (note that $\datav=\dictm\coefv+\errv$ can
be fully recovered from $(\errv,\coefv,\dictm)$),
\[ 
L(\datav,\coefv,\dictm) = L(\errv|\coefv,\dictm) + L(\coefv|\dictm) + L(\dictm).
\]
While computing each of these parts, three main issues need to be dealt
with:
\begin{enumerate}
\item {\bf Define appropriate probability models.}  Here, it is fundamental
  to incorporate as much prior information as possible, so that no cost is
  paid in learning (and thereby coding) already known statistical features
  of the data. Examples of such prior information include sparsity itself,
  invariance to certain transformations or symmetries, and (Markovian)
  dependencies between coefficients.
\item {\bf Deal with unknown parameters.}  We
  will use universal encoding strategies to encode data efficiently
  in terms of families of probability distributions. 
\item {\bf Model the effect of quantization.} All
  components, $\errv,\coefv,\dictm$ need to be quantized to some precisions,
  respectively $\delta_\err,\delta_\coef,\delta_\dict$, in order to obtain finite,
  realistic codelengths for describing $\datav$ (when the precision variable
  is obvious from the argument to which it is applied, we drop it to
  simplify the notation, for example, we will write
  $\quant{\err}_{\delta_\err}$ as $\quant{\err}$). This quantization introduces several
  complications, such as optimization over discrete domains, increase of
  sparsity by rounding to zero, increase of approximation error, and
  working with discrete probability distributions. \end{enumerate}
All such issues need to be considered with efficiency of computation in
mind. The discussion will focus first on the traditional, single-signal case
where each sample $\datav$ is encoded separately from the rest. At the end
of this section, we will also discuss the extension of this framework to a
multi-signal case, which has several algorithmic and modeling advantages
over the single-signal case, and which forms the basis for the dictionary
learning algorithms described later.

\subsection{Encoding the sparse coefficients}
\label{sec:encoding:coefficients}

\noindent {\bf Probability model:} Each coefficient in $\coefv$ is modeled
as the product of three (non-independent) random variables, $\coefrv =
\supprv\signrv(\valrv+\delta_\coef)$, where $\supprv=1$ implies $\coefrv
\neq 0$, $\signrv =\sgn(\coefrv)$, and
$\valrv=\max\{\abs{\coefrv}-\delta_\coef,0\}$ is the absolute value of
$\coefrv$ corrected for the fact that $\valrv\geq \delta_\coef$ when
$\supprv=1$. \footnote{Note that it is necessary to encode $\signrv$ and
  $\valrv$ separately, instead of considering $\signrv\valrv$ as one random
  variable, so that the sign of $\coefrv$ can be recovered when
  $|\valrv|=0$.}

\begin{figure*}\figspec
\begin{center}%
\includegraphics[height=1.5in]{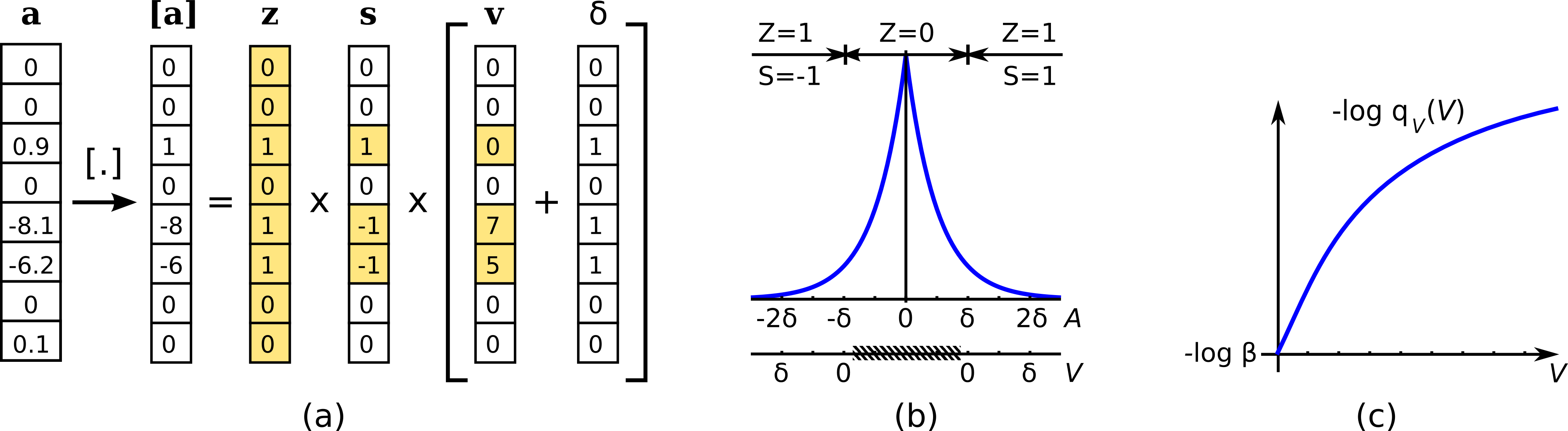}
\vspace{-0ex}\caption{\label{fig:coef-model} Encoding of the sparse
  code. (a) After quantization (here $\delta_\coef\!=\!1$), each coefficient
  $\coef_\ai$ is decomposed into three variables,
  $\supp_\ai\!=\!\Indicator(\coef_\ai)$,
  $\sign_\ai\!=\!\mathrm{sgn}(\coef_\ai)$ and
  $\val_\ai\!=\!\max\{|\coef_\ai|-\delta_\coef,0\}$. These are respectively
  modeled by random variables $\supprv \sim \Bernoulli(\rho_\coef)$,
  $\signrv \sim \Bernoulli(1/2)$, $\valrv \sim \Exponential(\theta_\coef)$
  (only the shaded numbers are actually encoded) (b) Scheme of the mapping
  from continuous coefficients (random variable $\coefrv$), into $\supprv$,
  $\signrv$ and $\valrv\xspace\!$. (c) Ideal codelength for the MOE model
  for $\valrv$, $-\log \cpdf{q}_\valrv(\valrv;\kappa,\beta)$. This is a
  smooth, concave function.}
\end{center}
\end{figure*}

We model $\supprv$ as a Bernoulli variable with
$\dpdf{p}(\supprv=1)=\rho_\coef$. Conditioned on $\supprv=0$,
$\signrv=\valrv=0$ with probability $1$, so no encoding is
needed.\footnote{We can easily extend the proposed model beyond
  $\signrv=\valrv=0$ and consider a distribution for $\signrv,\valrv$ when
  $\supprv=0$. This will naturally appear as part of the coding cost. This
  extends standard sparse coding to the case where the non-sparse component
  of the vector are not necessarily zero.} Conditioned on $\supprv=1$, we
assume $\dpdf{P}(\signrv=-1)=\dpdf{P}(\signrv=1)=1/2$, and $\valrv$ to be a
(discretized) exponential, $\Exponential(\theta_{\coef})$. With these
choices, $\dpdf{P}(\signrv\valrv|\supprv=1)$ is a (discretized) Laplacian
distribution, which is a standard model for transform (e.g., DCT, Wavelet)
coefficients. This encoding scheme is depicted in
Figure~\ref{fig:coef-model}(a,b). The resulting model is a particular case
of the ``spike and slab'' model used in statistics (see \cite{ishwaran05}
and references therein). A similar factorization of the sparse coefficients
is used in the Bayesian framework as well~\cite{carin11}.

\noindent {\bf Unknown parameters:} According to the above
model, the resulting encoding scheme for the coefficients (sparse code) is a three-parts
code: $L(\coefv) = L(\suppv) + L(\signv|\suppv) + L(\coefv|\signv,\suppv)$.
The support $\suppv$ is described using the \emph{enumerative two-parts
  code} \cite{cover73}, which first describes its size, $\norm{\coefv}_0$,
using $\log \natoms$ bits, and then the particular arrangement of the ones
in $\suppv$ using $\log {\natoms \choose \supsize}$ bits. The total
codelength for coding $\suppv$ is then
$
L(\suppv) = \log \natoms + \log {\natoms \choose \norm{\coefv}_0}.
$
This is a universal encoding scheme, and as such is more efficient than
those used previously in \cite{saito94,moulin99}. Then,
$L(\signv|\suppv)=\norm{\coefv}_0$ bits are needed to encode
$\signv\svec{\suppv}$, the actual signs of the non-zero
coefficients. Finally, we need to encode the magnitudes of the
$\norm{\coefv}_0$ non-zero coefficients, $\coefv\svec{\suppv}$. We do so by
considering it first as a sequence of exponentially-distributed continuous
random variables, to which quantization is applied later. Since the
parameter $\theta_\coef$ of the exponential is unknown,\footnote{This
  parameter is related to the sparsity level, and as discussed in
  Section~\ref{sec:background:issues}, is usually assumed known or
  determined via cross-validation. Following~\cite{ramirez10tip}, here we
  use tools from universal modeling, which permit to also automatically
  handle the non-stationarity of this parameter and its expected variability
  for different non-zero entries of $\coefv$.} we use a universal model
$\cpdf{q}_\valrv(\cdot)$ for the class of continuous exponential
distributions instead. We obtain such universal model
$\cpdf{q}_\valrv(\valrv)$ via a convex mixture, one of the standard techniques
for this,
\begin{align}
\cpdf{q}_\valrv(\valrv;\kappa_\coef,\beta_\coef) &=\!\!
  \int_{0}^{+\infty}{\!\!\!\!\!\!\!\!\Gamma(\theta;\kappa_\coef,\beta_\coef)\frac{\theta}{2}e^{-\theta|\valrv|}d\theta},
\label{eq:val-mixture}
\end{align}
where the mixing function
$\Gamma(\theta;\kappa,\beta)={\Gamma(\kappa)}^{-1}\theta^{\kappa-1}\beta^{\kappa}e^{-\beta\theta},$
is the Gamma density function of (non-informative) shape and scale parameters $\kappa$ and
$\beta$. With this choice, \refeq{eq:val-mixture} has a closed form
expression, and the degenerate cases $\theta=0$ and $\theta=\infty$ are
given zero weight. The resulting \emph{Mixture of Exponentials} (MOE)
density function $\cpdf{q}_\valrv(\valrv)$, is given by (see
\cite{ramirez10tip} for details),
\[
  \cpdf{q}_\valrv(\valrv;\beta_\coef,\kappa_\coef) = \kappa_\coef\beta_\coef^{\kappa_\coef}(\valrv+\beta_\coef)^{-(\kappa_\coef+1)},\;\valrv \in \reals^{+}.
\]
Note that the universality of this mixture model does not depend on the
values of the parameters $\kappa_\coef,\beta_\coef$, and guided by
\cite{ramirez10tip}, we set $\kappa_\coef=3.0$ and $\beta_\coef=50$. The
ideal Shannon codelength for this density function distribution is given by
$
-\log \cpdf{q}_\valrv(\valrv;\kappa_\coef,\beta_\coef) = -\log \kappa_\coef -\kappa_\coef\log \beta_\coef + (\kappa_\coef+1)\log(\valrv+\beta_\coef).$
This function, shown in Figure~\ref{fig:coef-model}(c), is non-convex,
however continuous and differentiable for $\valrv > 0$.

\noindent {\bf Quantization:} On one hand, quantizing the coefficients
to a finite precision $\delta_\coef$ increases the approximation/modeling error, from
$\datav-\dictm\coefv$ to $\datav-\dictm\quant{\coefv}_{\delta_\coef}$. This
additional error, $\dictm(\coefv-\quant{\coefv}_{\delta_\coef})$, will
clearly increase with $\delta_\coef$. On the other hand, larger
$\delta_\coef$ will reduce the description length of the non-zero values of
the coefficients, $\quant{\valv}_{\delta_\coef}$.
In practice, for reasonable quantization steps, the error added by such
quantization is negligible compared to the approximation error.  For
example, for describing natural images divided in patches of $8{\times}8$
pixels, our experiments indicate that there is no practical advantage in
using a value smaller than $\delta_\coef=16$. Consequently, our current
algorithms do not attempt to optimize the codelength on this parameter, and
we have kept this value fixed throughout all the experiments of
Section~\ref{sec:results}.


\subsection{Encoding the error}
\label{sec:encoding:error}

\begin{figure*}\figspec
\begin{center}
\includegraphics[width=0.95\textwidth]{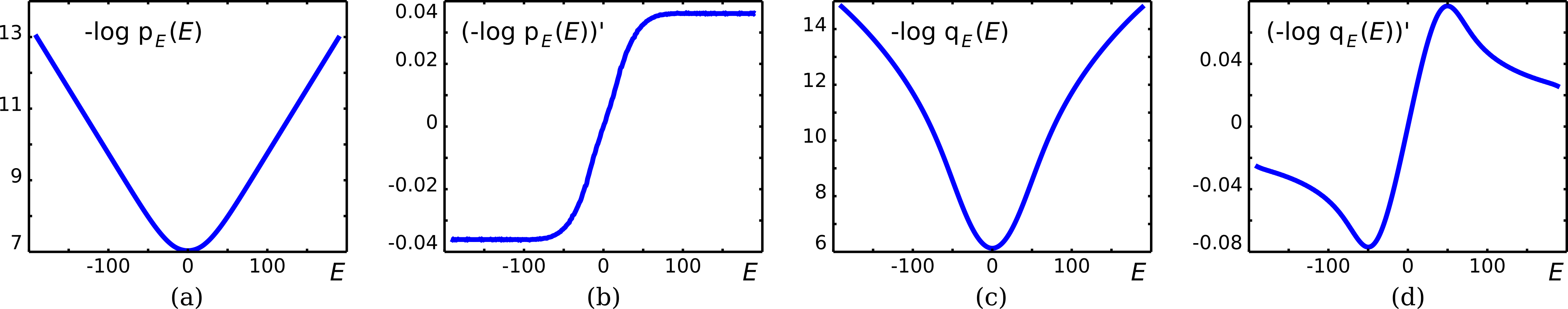}
\vspace{-0ex}\caption{\label{fig:error-model}Residual probability model. (a) Ideal
  codelength function of the ``LG'' distribution, $-\log
  \cpdf{p}_\errrv(\errrv)$, (b) LG influence function, that is, $(-\log
  \cpdf{p}_\errrv(\data))'$, (c) universal mixture for the LG model
  (MOEG), (d) MOEG influence function.}
\end{center}
\end{figure*}

\noindent {\bf Probability model:} Most sparse coding frameworks, including
all the mentioned MDL-based ones, assume the error $\errv$ to be solely due
to measurement noise, typically of the AWGN type. However, $\errv$ actually
contains a significant component which is due to a systematic deviation of
the model from the clean data. Following this, we model the elements of
$\errv$ as samples of an IID random variable $\errrv$ which is the linear
combination of two independent variables,
$\errrv=\hat{\errrv}+\noiserv$. Here $\noiserv \sim
\Gaussian(0,\sigma^2_\err)$ represents random measurement noise in $\datav$.
We assume the noise variance $\sigma^2_\err$ known, as it can be easily and
reliably estimated from the input data (for example, taking the minimum
empirical variance over a sufficient number of sub-samples). The
distribution of the second variable, $\hat{\errrv} \sim
\Laplacian(0,\theta_\err)$ is a Laplacian of unknown parameter
$\theta_\err$, which represents the error component due to the model itself.
The resulting continuous distribution $\cpdf{p}_\errrv(\errrv)$, which we
call ``LG,'' is the convolution of the distributions of both components (see
~\cite{ramirez10dude} for details on the derivation),
\begin{align}
\cpdf{p}_\errrv(\errrv;\sigma^2_\err,\theta_\err)  =& 
  \int_{\zeta=-\infty}^{+\infty}{\GaussianPDF[\sigma_\err]{\zeta}\LaplacianPDF[\theta_\err]{\errrv-\zeta}d\zeta} \nonumber \\
=& \frac{1}{4\theta_\err} e^{ \frac{\sigma^2_\err}{2\theta_\err^2} }
\left[
 e^{ \errrv/\theta_\err}\mathrm{erfc}\left( \frac{ \errrv+\sigma^2_\err/\theta_\err}{\sqrt{2}\sigma_\err} \right)+
 e^{-\errrv/\theta_\err}\mathrm{erfc}\left( \frac{-\errrv+\sigma^2_\err/\theta_\err}{\sqrt{2}\sigma_\err} \right)
\right],
\label{eq:lg-convolution}
\end{align}
where $\mathrm{erfc}(\aux) =
\frac{2}{\sqrt{\pi}}\int_{\aux}^{+\infty}{e^{-t^2}dt}$ is the
\emph{complimentary Gauss error function}. The ideal codelength, $-\log
\cpdf{p}_\errrv(\errrv)$, is shown in Figure~\ref{fig:error-model}(a) for
various parameter values. This function is convex and differentiable on
$\reals$, which is nice for optimization
purposes. Figure~\ref{fig:error-model}(b) shows its derivative, or so called
``influence function'' in robust statistics. It can be verified that $-\log
\cpdf{p}_\errrv(\errrv)$ behaves like a Laplacian with parameter $\theta_\err$
for large values of $\errrv$.  Further, since its derivative is bounded, the
influence of outliers is diminished. In fact, $-\log \cpdf{p}_\errrv(\errrv)$
is easily verified to be a $\psi$-type M-estimator, a family of functions
used in robust statistics (see \cite{huber64}).  Thus, using this model, we
obtain an information-theoretic robust estimator, which is consistent with
the motivations leading to its use in our framework, and which has a
significant practical impact in the experimental results.

\noindent {\bf Unknown parameters:} Since $\theta_\err$ is unknown, encoding
$\errv$ efficiently calls for the use of universal codes. In this case,
again, we employ a mixture model. Since the parameter $\theta_\err$ comes
from the underlying Laplacian component, we again use a Gamma for the mixing
function,
\begin{align}
\cpdf{Q}_\errrv(\errrv;\sigma^2_\err,\kappa_\err,\beta_\err) &= \!\! \int_{0}^{+\infty}{\!\!\!\!\!\!\!\!\Gamma(\theta;\kappa_\err,\beta_\err)\cpdf{p}_\errrv(\errrv;\sigma^2_\err,\theta)d\theta}.
\label{eq:lg-mixture}
\end{align}
%


We call this model MOEG. As with the MOE model, the universality of this
model is guaranteed by the theory for the choice of its underlying mixing
function, for any (non-informative) $\kappa_\err$ and $\beta_\err$. In this case, we use
$\kappa_\err=3.0$ and $\beta_\err=\delta_\err$. Also, we know from the
discussion above that $\sigma^2_\err$ can be easily and reliably estimated
from the data. Thus, we can say that the model for $\errrv$ is
parameter-free in this case as well. Figure~\ref{fig:error-model}(c) shows
the numerical evaluation of the ideal Shannon codelength $-\log
\cpdf{q}_\errrv(\errrv;\sigma^2_\err,\kappa_\err,\beta_\err)$, which is
non-convex.  However, it is twice differentiable
everywhere, again a desirable property for optimization purposes (more on
this in sections~\ref{sec:encoding-algorithms}
and~\ref{sec:learning-algorithms}). As with the LG distribution, $-\log
\cpdf{q}_\errrv(\errrv)$ is an $\psi$-type M-estimator, in this case, a
\emph{redescending} M-estimator, since its derivative
(Figure~\ref{fig:error-model}(d)) vanishes to $0$ at $\infty$. As such,
$-\log \cpdf{q}_\errrv(\errrv)$, derived from the universal model
corresponding to $\cpdf{p}_\errrv(\errrv)$, can reject outliers even more
aggressively than $-\log \cpdf{p}_\errrv(\errrv)$, again marrying robust
statistics with information theory in a natural way.

\noindent {\bf Quantization:} To losslessly encode finite-precision input
data such as digital images, the quantization step of the error
coefficients needs not be more than that of the data itself,
$\delta_\data$, and we simply quantize the error coefficients uniformly with
step $\delta_\err=\delta_\data$. For example, for $8$-bit digital images, we
set $\delta_\err=\delta_\data=1$.
%

\subsection{Model for the dictionary}
\label{sec:encoding:dictionary}
\noindent {\bf Probability model:} Dictionary learning practice shows that
learned atoms, unsurprisingly, present features that are similar to
those of the original data.  For example, the piecewise smoothness 
of small image patches is to be expected in the atoms of learned
dictionaries for such data. This prior information, often neglected in dictionary learning algorithms, needs to be taken into
account for encoding such atoms efficiently.

\begin{figure*}\figspec
\begin{center}
\includegraphics[width=0.95\textwidth]{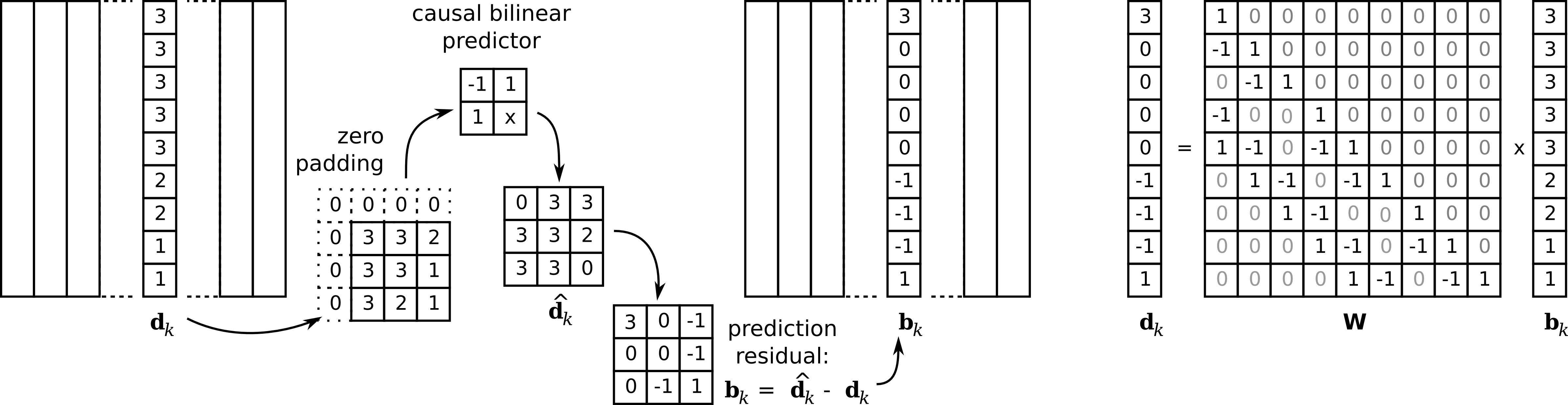}%
\caption{\label{fig:dict-model} Prediction scheme used for learning natural
  image patches dictionaries (in this example, $3{\times}3$ patches, and
  $\ndims=9$). An atom $\dictv_\ai$ is arranged as a ${3{\times}3}$ patch,
  and a causal bi-linear predictor (shown as a $2{\times}2$ template) with
  zero-padding (pixels outside of the patch are assumed $0$) is applied to
  it, producing a predicted atom $\hat\dictv_\ai$ and a residual
  $\dpredv_\ai=\dictv_\ai-\hat\dictv_\ai$. The previous operation can be
  written as $\dpredv_\ai=\mat{W}\dictv_\ai$, with $\mat{W} \in
  \reals^{9{\times}9}$ the linear mapping from atom to prediction
  residuals corresponding to this example.}
\end{center}
\end{figure*}

We embody such information in the form of \emph{predictability}. This is, we
will encode an atom $\dictv \in \datavspace$ as a sequence of causal
prediction residuals, $\dpredv \in \datavspace$, $b_{i+1} =
\dict_{i+1}-\tilde\dict_{i+1}(\dict_1,\dict_2,\ldots,\dict_i),\, 1 \leq i <
\ndims$, a function of the previously encoded elements in $\dictv$. In
particular, if we restrict $\tilde\dict_{i+1}$ to be a linear function, the
residual vector can be written as $\dpredv = \mat{W}\dictv$, where $\mat{W}
\in \reals^{\ndims{\times}\ndims}$ is lower triangular due to the causality
constraint (this aspect has important efficiency consequences in the
algorithms to be developed in Section~\ref{sec:learning-algorithms}).  This
is depicted in Figure~\ref{fig:dict-model}, along with the specific
prediction scheme that we adopted for the image processing examples in
Section~\ref{sec:experiments}. In this case we consider an atom $\dictv$ to
be an $\sqrt{\ndims}{\times}\sqrt{\ndims}$ image patch, and use a causal
bi-linear predictor where the prediction of each pixel in the dictionary
atom is given by $\mathrm{north\_pixel} + \mathrm{west\_pixel} -
\mathrm{northwest\_pixel}$.

As a general model for linear prediction residuals, we assume $\dpredv$ to
be a sequence of IID Laplacian samples of parameter $\theta_\dict$.  In
principle, $\theta_\dict$ is also unknown.  However, describing $\dictm$ is
only meaningful for dictionary learning purposes, and, in that case,
$\dictm$ is updated iteratively, so that when computing an iterate
$\dictm\iter{t}$ of $\dictm$, we can use $\dictm\iter{t-1}$ to estimate and
fix $\theta_\dict$ via ML (more on this $\theta_\dict$ later in
Section~\ref{sec:learning-algorithms}). Thus, we consider $\theta_\dict$ to
be known.

\noindent {\bf Quantization:} When $\coefm$ is fixed during a dictionary
learning iteration (which consists of an alternate descent between $\dictm$
and $\coefm$), we can view $(\coefm,\datam)$ as $\nsamples$ input-output
training pairs, and $\dictm$ as the ML estimator of the linear coefficients
describing such mapping via $\datam=\dictm\coefm+\errm$. Based on this, we
use the quantization step $\delta_\dict=1/\sqrt{\nsamples}$, which is an
optimal step for encoding the ML parameter in two-part codes, as described
in \cite[Theorem~1]{rissanen84}.

\noindent {\bf Computation:} Computing $L(\dictm)$ is only relevant for
learning purposes. In general, since $\norm{\dictv_k}_2 \leq 1$, and
$\norm{\dictv_k}_2 \leq \sqrt{\ndims}\norm{\dictv_k}_1$, we have that
$\hat\theta_\dict=(\natoms\ndims)\inv\sum_\ai\norm{\dictv_\ai }_1 \leq
(\natoms\sqrt{\ndims})\inv \ll \delta_\dict=\sqrt{\nsamples}$, and the error
of using the approximation \refeq{eq:approx-codelength} is not significant,
\begin{align}
L(\dictm) =& \sum_{\ai=1}^{\natoms} L(\dictv_\ai) \approx \sum_{\ai=1}^{\natoms}\left\{ -\log \cpdf{p}(\mat{W}\dictv_\ai;\theta_\dict) - \ndims\log \delta_\dict \right\}
=  \theta_\dict\sum_{\ai=1}^{\natoms}{\norm{\mat{W}\dictv_\ai}_1} + \frac{\ndims\natoms}{2}\log \nsamples + c,
\end{align}
where $\cpdf{p}(\mat{W}\dictv_\ai)$ is the IID Laplacian distribution over
the $\ai-$th atom prediction residual vector $\mat{W}\dictv_\ai$, and $c$
is a fixed constant. For $\natoms$ fixed (we will later see how to learn the
dictionary size $\natoms$ as well), the above expression is simply an
\cost{1} penalty on the atom prediction residual coefficients. As we will
see in Section~\ref{sec:learning-algorithms}, this allows us to use
efficient convex optimization tools to update the atoms.

\subsection{Extension to sequential (collaborative) coding}
\label{sec:encoding:collaborative}

One natural assumption that we can often make on the set of data samples
$\datam$ is that, besides all being sparsely representable in terms of the
learned dictionary $\dictm$, they share other statistical properties. For example,
we can assume that the underlying unknown model parameters, $\theta_\err$,
$\rho_\coef$, $\theta_\coef$, $\theta_\dict$, are the same for all columns of the
sparse data decomposition ($\errm$, $\coefm$).  

\begin{figure*}\figspec
\begin{center}
\includegraphics[width=\textwidth]{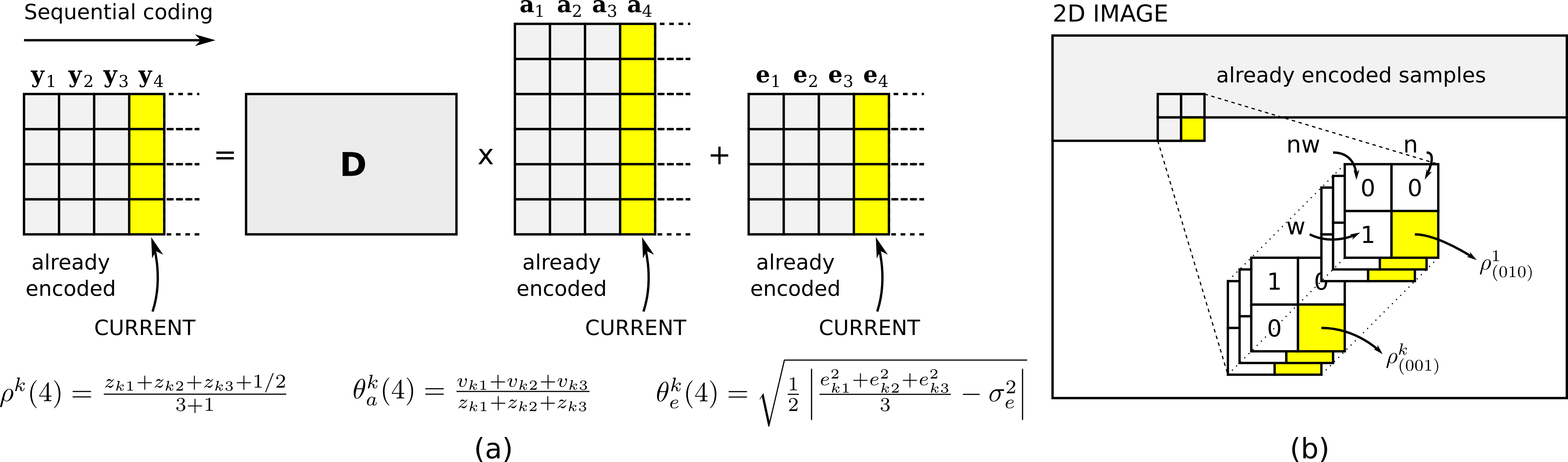}%
\caption{\label{fig:collaborative-encoding}Collaborative encoding
  scheme. (a) In this example, $3$ samples have already been encoded, and we
  are about to encode sample $4$. The formulas for estimating the various
  model parameters are shown for $j=4$, in particular those for the error
  and the coefficients associated to the $\ai$-th atom (the $\ai$-th row of
  $\coefm$). (b) Markov model for the coefficients support matrix
  $\suppm$. Here, a sample patch $\datav$ is about to be encoded. Here the
  first atom was only used by the pixel to the west, so that the Markov
  state for modeling $\supp_1$ is $(n,w,nw)=(0,1,0)$, and
  $P(\supp_1=1)=\rho_{(0,1,0)}^{1}$. As for the $\ai$-th atom, only the $nw$
  pixel has used it, so that the Markov state for $\supp_\ai$ is $(0,0,1)$,
  that is, $P(\supp_\ai=1)=\rho_{(0,0,1)}^{\ai}$. }
\end{center}
\end{figure*}

Under such assumption, if we encode each column of $\datam$ sequentially%
, we can learn statistical information from the ones already encoded and
apply it to estimate the unknown parameters of the distributions used for
encoding the following ones. The general idea is depicted in
Figure~\ref{fig:collaborative-encoding}(a).  Concretely, suppose we have
already encoded $j-1$ samples. We can then use
$[\errv_1,\errv_2,\ldots,\errv_{(j-1)}]$ to estimate $\theta_\err$, and
$[\coefv_1,\coefv_2,\ldots,\coefv_{(j-1)}]$ to estimate $\theta_\coef$ and
$\rho_\coef$, and ``plug-in'' these parameters to encode the $\si$-th
sample. This justifies the name of this encoding method, which is known in
the coding literature as \emph{sequential plug-in} encoding. This encoding
strategy has several advantages: 1) For common parameter estimators such as
ML, this method can be shown to be universal; 2) Since all distribution
parameters are fixed (pre-estimated) when encoding the $\si$-th sample, we
can use the ``original,'' non-universal distributions assumed for modeling
$\errv_j$ (LG) and $\coefv_j$ (Laplacian), which have closed forms and are
usually faster to compute (together with \refeq{eq:approx-codelength}) than
their universal mixture counterparts; 3) Furthermore, these original
distributions are convex, so that in this case, given a fixed support, we
are able to exactly minimize the codelength over the non-zero coefficient
values; 4) With many samples available for parameter estimation, we can
potentially afford more complex models. 

\noindent{\bf Residual:} We estimate $\theta_\err$ in two steps. First,
since the random variable $\errrv$ is an independent sum of two random
variables, $\errrv=\hat{\errrv}+\noiserv$, we have that
$\fun{var}(\errrv)=\fun{var}(\hat\errrv)+
\fun{var}(\noiserv)=\fun{var}(\hat{\errrv})+\sigma_\err^2$. Now, since
$\hat{\errrv}$ is Laplacian, we have that
$\fun{var}(\hat{\errrv})=2\theta_\err^2$. Combining both equations we have
that $\theta_\err=0.5\sqrt{\fun{var}(\hat\errrv)-\sigma_\err^2}$.  With the
noise variance $\sigma_\err^2$ assumed known, and using the standard
unbiased variance estimator,
$\hat{\fun{var}}(\hat\errrv)=(\natoms(\si-1))^{-1}\norm{\errm\svec{1,\ldots,(\si-1)}}_F^2$,
we obtain
\[
\hat\theta_\err=0.5\sqrt{\mathrm{max}\{(\natoms(\si-1))^{-1}\norm{\errm\svec{1,\ldots,(\si-1)}}_F^2-\sigma_\err^2,0\}},
\]
where the maximization guarantees that $\hat\theta_\err \in \reals^+$.

\noindent{\bf Coefficients:} In the case of $\coefv$, we have in principle
two unknown parameters, the probability of an element being non-zero,
$\rho_\coef$, and the scale parameter of the Laplacian governing the
non-zero values, $\theta_\coef$ (both previously handled with universal
models). Here, however, we extend the model, drawing from the well known
empirical fact that coefficients associated to different atoms can have very
different statistics, both in frequency and variance. This is typical of DCT
coefficients for example (see \cite{lam00}), and has been consistently
observed for learned dictionaries as well~\cite{ramirez10tip}. Therefore, we
will consider a separate set of parameters
$(\rho_\coef^\ai,\theta_\coef^\ai)$ for each \emph{row} $\ai$ of $\coefm$,
$\coefv^\ai$. We update such parameters from the coefficients observed in
the respective row for the already-computed samples,
$(\coef_{k1},\coef_{k2},\ldots,\coef_{k(j-1)})$, and encode each $\ai$-th
coefficient in $\coefv_j$ (more specifically, in $\suppv_j$, and $\valv_j$),
as the $\si$-th sample of the respective row.  Concretely, let $n_1^\ai =
\sum_{\si'=1}^{(\si-1)}{\supp_{\ai\si'}}$ be the number of non-zero
coefficients observed so far in the $\ai$-th row. For $\rho_\coef^\ai$, we
use the Krichevsky-Trofimov (KT) estimator~\cite{krichevsky81},
\begin{equation}
\hat\rho_\coef^\ai = \frac{n_1^\ai + 0.5}{j},
\label{eq:kt}
\end{equation}
which is a universal plug-in encoding scheme for Bernoulli sequences of
unknown parameter. For encoding $\val_{\ai\si}$, we apply the ML estimator for
the exponential family to the non-zero coefficients observed so far in the
$\ai$-th row. Recalling that
$\val_{\ai\si}=\max\{|\coef_{\ai\si'}|-\delta_\coef,0\}$, the resulting
estimator is given by
\[ \hat\theta_\coef^\ai = \frac{\sum_{\si'=1}^{(\si-1)} \max\{|\coef_{\ai\si'}|-\delta_\coef,0\}}{n_1^\ai }.
\]

\noindent {\bf Markovian dependencies:} In many applications,
spatially/temporally adjacent samples are statistically dependent. For
example, we may assume that an atom is more likely to occur for a sample
$\si$ if it has been used by, say, the $(\si-1)$-th sample (see also
\cite{zhou11aistats}). In that case, we may consider different estimations
of $\rho^\ai$ depending on the value of $\supp_{\ai(\si-1)}$, $\rho^\ai_1 =
\dpdf{P}(\supp_{\ai\si}=1|\supp_{\ai(\si-1)}=1)$, and $\rho^\ai_0 =
\dpdf{P}(\supp_{\ai\si}=1|\supp_{\ai(\si-1)}=0)$.  In particular, for the
image processing results of Section~\ref{sec:results}, we use a Markovian
model which depends on three previous samples, corresponding to the
(causal) neighboring west, north, and northwest patches of the one being
encoded. Thus, for each atom $\ai$ we will have $8$ possible parameters,
$\rho^\ai_{(n,w,nw)}, (n,w,nw) \in \setdef{0,1}^3$, where each value of
$(n,w,nw)$ indicates a possible \emph{Markov state} in which a sample may
occur. This is depicted in Figure~\ref{fig:collaborative-encoding}(b). For
each state $(n,w,nw)$, we estimate $\rho^\ai_{(n,w,nw)}$ using \refeq{eq:kt}, 
with the average taken over the samples which occur in the same state $(n,w,nw)$.

\section{MDL based sparse coding}
\label{sec:encoding-algorithms}

For the encoding problem, $\dictm$ is fixed (it has already been learned),
and we consider encoding a single data sample $\datav$. The model selection
problem here is that of choosing the model (indexed by the sparse code
$\coefv$) among all the models belonging to the nested family of model
classes $\modelclass(\gamma) =\setdef{\coefv \in \coefvspace,
  \norm{\coefv}_0 \leq \gamma}, \gamma=0,\ldots,\natoms$, that yields the
smallest codelength for describing $\datav$. In principle, this calls for
finding the best model $\coefv(\gamma)$ within each model class
$\modelclass(\gamma)$, and then selecting $\opt\coefv = \arg\min_{0\leq
  \gamma \leq \natoms} L(\datav,\coefv(\gamma))$. However, in order to be
computationally efficient, and as with most sparse coding and model
selection algorithms, several simplifications and approximations are needed.
Let us first consider the problem of finding $\coefv(\gamma)$,
\begin{align}
\coefv(\gamma) &\defeq \arg\min_{\coefv \in \modelclass(\gamma)}
L(\datav,\coefv) \nonumber
= \arg\min_{\coefv \in \coefvspace} -\log \dpdf{p}_\errrv(\datav-\dictm\coefv)  -\log\dpdf{p}(\suppv) -\log\dpdf{p}(\signv|\suppv) 
- \log \dpdf{p}_\valrv(\coefv|\signv,\suppv) \nonumber \\
&= \arg\min_{\coefv \in \coefvspace} -\log \dpdf{p}_\errrv(\datav-\dictm\coefv) - \log {\natoms \choose \norm{\coefv}_0} + \norm{\coefv}_0 -
\log \dpdf{p}_\valrv(\coefv) \st \norm{\coefv}_0 \leq \gamma.
\label{eq:fixed-support}
\end{align}
For quantized $\coefv$, this is an optimization problem over a discrete,
infinite domain, with a non-convex (in the continuous domain) constraint,
and a non-differentiable cost function in $\coefv$.  Based on the literature
on sparse coding, at least two alternatives can be devised at this
point. One way is to use a pursuit technique, e.g.,
~\cite{mallat93}. Another option is to use a convex relaxation of the
codelength function, e.g., ~\cite{chen98}. For the sake of brevity, here we
will describe an algorithm loosely based on the first alternative. Details
on the convex relaxation method for MDL-based sparse coding will be
published elsewhere.

The pursuit-like algorithm, which we call COdelength-Minimizing Pursuit
Algorithm (COMPA), is summarized in Algorithm~\ref{alg:compa}.  This is a
non-greedy cross-breed between Matching Pursuit (MP) and Forward Stepwise
Selection (FSS)~\cite{hastie09}. As with those methods, COMPA starts with
the empty solution $\coefv\iter{0}=\vec{0}$, and updates the value of one
single coefficient at each iteration. Then, given the current correlation
$\corrv\iter{t}=\dictm\errv\iter{t}$ between the dictionary
atoms and the current residual, each $\ai$-th coefficient in
$\coefv\iter{t}$ is tentatively incremented (or decremented) by
$\Delta_\ai=\quant{\corr\iter{t}_\ai}$, and a candidate codelength $
\hat L_\ai$ is computed in each case. The coefficient that
produces the smallest $\hat L(\datav,\coefv)$ is updated to produce
$\coefv\iter{t+1}$.

The logic behind this procedure is that the codelength cost of adding a new
coefficient to the support is usually very high, so that adding a new
coefficient only makes sense if its contribution is high enough to produce
some noticeable effect in the other parts of the codelength. A variant of
this algorithm was also implemented where, for each candidate $\ai$, the
value of the increment $\Delta_\ai$ was refined in order to minimize $
\hat L_\ai$. However, this variant turned out to be significantly
slower, and the compression gains where below $0.01$ bits per sample
(uncompressed codelength is $8$ bits per sample). Assuming that $L\iter{t}$
is unimodal, the algorithm stops if the codelength of a new iterate is
larger than the previous one. To assess the validity of this assumption, we
also implemented a variant which stops, as MP or FSS, when the
residual-coefficients correlation $\norm{\corrv\iter{t}}_\infty$ is no
longer significant, which typically requires many more iterations. With this
variant we obtained a negligible improvement of $0.004$ bits per sample,
while increasing the computational cost about three times due to the extra
iterations required.

\begin{figure*}\figspec
\begin{center}
\includegraphics[width=0.7\textwidth]{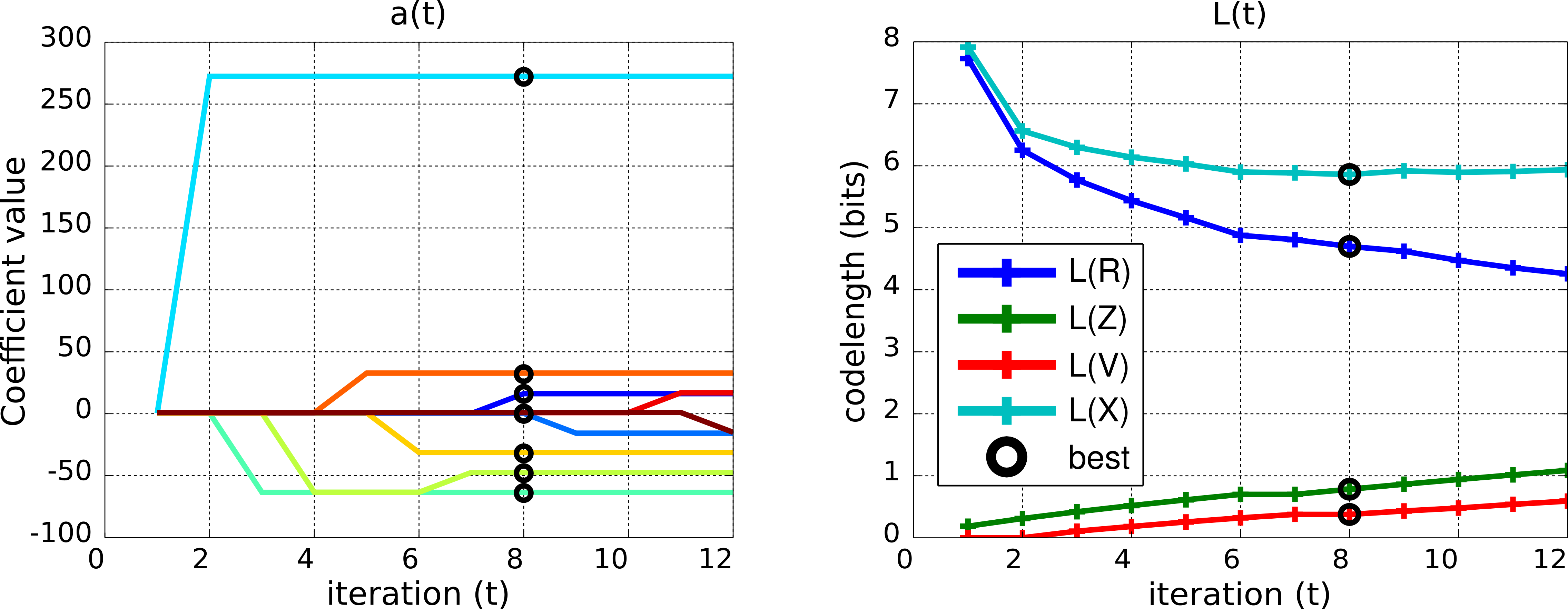}
\caption{\label{fig:compa-evolution}Typical evolution of the COMPA
  algorithm. (a) coefficients. (b) codelength. The best iterate (code) is
  marked with a black circle. Also note that describing the support ($L(Z)$)
  actually takes more bits than describing the non-zero values ($L(V)$).  }
\end{center}
\end{figure*}
\begin{algorithm}[t]
\begin{scriptsize}
  \caption{\label{alg:compa}COdelength Minimizing Pursuit Algorithm (COMPA)}
\SetKw{Init}{initialize}
\SetKw{Set}{set}
\SetKw{Choose}{choose}
\SetCommentSty{textit}
\KwIn{Data sample $\datav$, dictionary $\dictm$}
\KwOut{$\opt\coefv$, $\opt\errv$}
\Init $t \assign 0; \coefv\iter{0} \assign \mat{0}; \errv \assign \datav;  L\iter{0} \assign L(\datav,\mat{0});\; \corrv\iter{t} \assign \dictm^T\errv\iter{t}$ \tcp*{$\corrv\iter{t}$ correlation of current residual with the dictionary}
\Repeat{$L\iter{t} \geq L\iter{t-1}$}{
  \For{$\ai \assign 1,2,\ldots,\natoms $}{
    $\Delta_{\ai} \assign [\corr\iter{t}_\ai]_{\delta_\coef}$  \tcp*{step $\Delta_{\ai}$ is correlation, quantized to prec. $\delta_\coef$}
    $\tilde{L}_{\ai} \assign L([\errv-\Delta_\ai\dictv_\ai]_{\delta_\err},\,\coefv + \Delta_\ai \canon_\ai)$ \tcp*{$\canon_\ai\!=$ $\ai$-th canonical vec. of $\reals^{\natoms}$}

  }
  $L\iter{t+1} \assign \min \{ \tilde{L}_{\ai}: \ai=1,\ldots,\natoms\}$ \;
  $\coefv\iter{t+1} \assign \coefv\iter{t} + \Delta_{\opt{\ai}} \canon_{\opt{\ai}}$ 
  \tcp*{update coefficients vector}
  $\corrv\iter{t+1} \assign \corrv\iter{t} - \Delta_{\opt{\ai}} \dictv_{\opt{\ai}}$
  \tcp*{update correlation}
  $t \assign t + 1$ \;
}
$\opt\coefv \assign \coefv(\gamma-1)$ \;
$\opt\errv \assign \quant{\datav-\dictm\opt\coefv}$ \;
STOP \;
\end{scriptsize}
\end{algorithm}

\section{MDL based dictionary learning}
\label{sec:learning-algorithms}

Given that our sparse coding algorithm in
Section~\ref{sec:encoding-algorithms} can select the best support size
$\gamma$ for each sample in $\datam$, the definition of the model class
$\modelclass(\maxnz,\natoms)$ given in
Section~\ref{sec:mdl-model-selection}, which assumes the same $\gamma$ for
all samples in $\datam$, is no longer appropriate (we could of course add
$0$-weight coefficients to make $\gamma$ equal for all data). Instead, for
dictionary learning, we consider the model class family
$\modelclass(\natoms) = \setdef{(\coefm,\dictm),\dictm \in \dictmspace,
  \coefv_\si \in \modelclass(\gamma;\dictm), \si=1,\ldots,\nsamples}$, where
$\modelclass(\gamma;\dictm)$ is the model class family of sparse codes based
on a fixed dictionary $\dictm$ defined in
Section~\ref{sec:encoding-algorithms}, with the dependency on $\dictm$ made
explicit. It is easy to see that the model classes $\modelclass(\natoms)$
are nested. We now need to solve
\begin{equation}
(\coefm(\natoms),\dictm(\natoms)) =
\arg\min_{(\coefm,\dictm) \in \modelclass(\natoms)} L(\errm,\coefm,\dictm),
\label{eq:dictionary-learning-problem}
\end{equation}
 for $\natoms\!=\!0,1,\ldots$, and then choose
 $(\opt\coefm,\opt\dictm)=(\coefm(\hat{\natoms}),\dictm(\hat{\natoms}))$
 with the optimal dictionary size $$\hat{\natoms} = \arg\min_\natoms
 \setdef{ L(\errm,\coefm(\natoms),\dictm(\natoms)):
   \natoms\!=\!0,1,\ldots}. $$ As with sparse coding, here we exploit the
 nested nature of the model classes to speed up the model selection. For
 this, we propose a forward-selection algorithm, described in
 Algorithm~\ref{alg:mdl-dl-forward}, which starts from $\modelclass(0)$ (the
 empty dictionary), and then approximates the best model in
 $\modelclass(\natoms+1)$ by adding a new atom to the dictionary computed
 for $\modelclass(\natoms)$ and then invoking Algorithm
 ~\ref{alg:mdl-dictionary-learning}, which is discussed in depth in the next
 subsection.

A backward-selection algorithm was also developed which first learns the
model for $\modelclass(\natoms_{\mathrm{max}})$ via 
\refeq{alg:mdl-dictionary-learning}, where $\natoms_{\mathrm{max}}$ is a
given maximum dictionary size, and then prunes the less frequently used
atoms until no further decrease in codelength is observed. This algorithm
allows us to provide especially-constructed initial dictionaries for
Algorithm~\refeq{alg:mdl-dictionary-learning}, e.g., an (overcomplete) DCT
frame, which can be critical for finding good local minima of the non-convex
problem \refeq{eq:dictionary-learning-problem}. We do this for example to
learn a dictionary for the whole class of natural images, see
Section~\ref{sec:experiments}.


\begin{algorithm}[t]
\begin{scriptsize}
\caption{\label{alg:mdl-dl-forward}MDL-based dictionary learning via forward
  selection.}  \SetKw{Init}{initialize} \SetKw{Set}{set}
\SetKw{Choose}{choose} \SetCommentSty{textit} \KwIn{Data $\datam$}
\KwOut{$(\opt{\coefm},\opt{\dictm})$} \Init $\natoms\assign0$;
$\coefm(0)\assign\emptyset$; $\dictm(0)\assign\emptyset$; $\errm(0) \assign
\datam$; $L(0) \assign L(\errm(0),\coefm(0),\dictm(0))$ \;
\Repeat{$L(\natoms) \geq L(\natoms-1)$ } { $\tilde\dictv \assign \vec{u}_1,
  \mat{U}\Sigma\mat{V}\transp = \errm(\natoms)$ \tcp{Initial value of new
    atom is the left-eigenvector associated to the largest singular value of
    $\errm\iter{t}$.}  $\dictm^0 \assign
      [\,\dictm(\natoms)\,|\,\tilde\dictv\,]$ \tcp{Initial dictionary for
        optimization below.}  $(\coefm(\natoms+1),\dictm(\natoms+1)) \assign
      \arg\min_{(\coefm,\dictm) \in \modelclass(\natoms+1)}
      L(\errm,\coefm,\dictm)$ \tcp{Optimize dict. via
        Algorithm~\ref{alg:mdl-dictionary-learning}} $ \natoms \assign
      \natoms + 1 $ \; $L(\natoms) \assign
      L(\errm(\natoms),\coefm(\natoms),\dictm(\natoms))$ \; } $\opt\coefm
\assign \coefm(\natoms-1)$; $\opt\dictm \assign \dictm(\natoms-1)$\;
\end{scriptsize}
\end{algorithm}

\subsection{Optimizing the dictionary for fixed p}

For fixed $\natoms$, and given an initial $\dictm$,
Algorithm~\ref{alg:mdl-dictionary-learning} adapts the atoms of $\dictm$ to
fit the training data $\datam$. 
\begin{algorithm}[t]
\begin{scriptsize}
  \caption{MDL-based dictionary learning for a given size $\natoms$}
\label{alg:mdl-dictionary-learning}
\SetKw{Init}{initialize}
\SetKw{Set}{set}
\SetKw{Choose}{choose}
\SetCommentSty{textit}
\KwIn{Data $\datam$, initial dictionary $\dictm^0$, multiplier $\lambda$, $\eta$}
\KwOut{Local-optimum $(\opt{\coefm},\opt{\dictm})$}
\Init $\dictm\iter{0}=\dictm^0$, $t=1$ \;
\Repeat{$ \frac{ \norm{ \dictm\iter{t} - \dictm\iter{t-1} }_2 }{ \norm{ \dictm\iter{t} }_2 } \leq \mathrm{\epsilon}$ }
{
  \For{$j = 1,\ldots,\nsamples$ } {
    $\coefv_j\iter{t} \assign \arg\min_{\coefm} L(\errv,\coefv,\dictm\iter{t-1})$ \;
  }
  Update plug-in parameters: $\theta_\err, \setdef{(\theta_\coef^\ai,\rho_\coef^\ai),\ai=1,\ldots,\natoms}, \theta_\dict$\;
  $\dictm\iter{t} \assign \arg\min_{\dictm} L(\errm,\coefm,\dictm)$ \;
  $ t \leftarrow t + 1 $ \;
}
\end{scriptsize}
\end{algorithm}
At the high level, our algorithm is very similar to the traditional approach
of alternate minimization over $(\coefm,\dictm)$. However, there are a
number of important differences, namely: 1) The cost function minimized is now
the cumulative codelength of describing $\datam$,
$L(\errm,\coefm,\dictm)$; 2) Minimizing over $\coefm$ is done sample by sample
following Section~\ref{sec:encoding-algorithms}; 3) Since $\dictm$ needs to be
described as well, it has an associated codelength (see
Section~\ref{sec:encoding:dictionary}), resulting in regularized dictionary
update, described below; 4) in a cross-breed between Expectation-Maximization,
and plug-in estimation, we estimate the model parameters for the current
iterate $(\errm\iter{t},\coefm\iter{t},\dictm\iter{t})$, from the
accumulated statistics of previous iterates
$\setdef{(\errm\iter{t'}\coefm\iter{t'},\dictm\iter{t'}),t'=1,\ldots,t-1}$.
At the end of the learning process, these parameters are ``saved'' as part
of the learned model and can be used for modeling future data along with
$\dictm$.

At the $t$-th iteration of the alternate minimization between $\dictm$ and
$\coefm$, with $\coefm\iter{t}$ just computed and kept fixed, the dictionary
step consists of solving the sub-problem $$\dictm\iter{t} = \arg\min_{\dictm
  \in \dictmspace} L(\datam,\coefm\iter{t},\dictm) = \arg\min_{\dictm
  \in \dictmspace} L(\datam|\coefm\iter{t},\dictm)+L(\dictm).$$  According to
Section~\ref{sec:encoding:dictionary}, we have
$L(\dictm)=\frac{1}{\theta_\dict\iter{t}}\sum_{\ai=1}^{\natoms}
\norm{\mat{W}\dictv_\ai}_1$, where
$\theta_\dict\iter{t}=\frac{1}{\ndims\natoms}\sum_{\ai=1}^{\natoms}\sum_{\di=1}^{\ndims}|\dict_{\di\ai}\iter{t-1}|$
is the Laplacian MLE of $\theta_\dict$ based on $\dictm\iter{t-1}$.
Correspondingly, the data fitting term, via~\refeq{eq:approx-codelength} and
disregarding the constant terms, is given by
$L(\datam|\coefm\iter{t},\dictm) =
L(\datam-\dictm\coefm\iter{t}|\theta_\err\iter{t},\sigma_\err^2) =
\sum_{\si=1}^{\nsamples}\sum_{\di=1}^{\ndims} -\log
LG(\data_{\di\si}-(\dictm\coefm\iter{t})_{\di\si};\theta_\err\iter{t},\sigma_\err^2)$,
where $\theta_\err\iter{t}$ is the estimator of $\theta_\err$ given
$\errm=\datam-\dictm\iter{t-1}\coefm\iter{t}$ (see Section
\ref{sec:encoding:collaborative}) and $\sigma_\err^2$ is assumed known.  The
problem can now be written as,
\begin{equation}
  \dictm\iter{t} = \arg\min_{\dictm} L(\datam-\dictm\coefm\iter{t}|\theta_\err\iter{t},\sigma_\err^2) +  \theta_\dict\iter{t}\sum_{\ai=1}^{\natoms} \norm{\mat{W}\dictv_\ai}_1.
\label{eq:learning-cost-1}
\end{equation}
For general $\mat{W}$, the optimization of \refeq{eq:learning-cost-1} is
challenging since none of the above terms are separable, in particular, the
non-differentiable \cost{1} term. However, since $\mat{W}$ is easily
invertible (as described in
Section~\ref{sec:encoding:dictionary}, it is lower triangular with $1$'s in
the diagonal), we can perform a change of variables and solve the equivalent
problem in the \emph{prediction residual matrix} $\mat{U}=\mat{W}\dictm$ instead,
\begin{equation}
  \opt{\mat{U}} = \arg\min_{\mat{U}} L(\datam-\mat{W}\inv\mat{U}\coefm\iter{t}|\theta_\err\iter{t},\sigma_\err^2) +  \theta_\dict\iter{t}\sum_{k=1}^{\natoms} \norm{\mat{u}_k}_1.
\label{eq:learning-cost-2}
\end{equation} 
Since the regularization term in~\refeq{eq:learning-cost-2} is decoupled in
the elements of $\mat{U}$, and
$L(\datam-\mat{W}\inv\mat{U}\coefm|\theta_\err\iter{t},\sigma_\err^2)$ is
convex and differentiable in $\mat{U}$ (see
Figure~\ref{fig:error-model}(a)), \refeq{eq:learning-cost-2} can be
efficiently solved using the existing techniques for separable
non-differentiable regularization terms. In our case, we employ the
backtracking variant of FISTA~\cite{beck09siam}, focusing on an efficient
numerical evaluation of each step.

\section{Experimental results}
\label{sec:results}
\label{sec:experiments}

\subsection{Coding performance}
\label{sec:results:coding}

The first experiment in this section assesses the ability of our coding
scheme to actually produce a compressed description of the data, in this
case $8$-bit gray-scale images. To this end, a dictionary $\dictm$ was
learned using the backward-selection algorithm, for the training samples
from the Pascal'06 image
database, \footnote{\small\url{http://pascallin.ecs.soton.ac.uk/challenges/VOC/databases.html}}
converted to $8$-bit gray-scale images and decomposed into $8{\times}8$
patches. The initial dictionary was an overcomplete DCT frame with
$\natoms=256$. The resulting global dictionary $\dictm$ has $\natoms=250$
atoms. We then encoded the testing samples from the same database, obtaining
an average codelength of $4.1$ bits per pixel (bpp), confirming the ability of
our model to produce a compressed description of the data.

\subsection{Learning performance}
\label{sec:results:learning}

We compare the performance of the forward and backward dictionary learning
algorithms proposed in Section~\ref{sec:learning-algorithms} by applying
each method to learn a dictionary for the standard ``Boats'' image (taken
from the SIPI
database, \footnote{\small\url{http://sipi.usc.edu/database/database.php?volume=misc\&image=38\#top}}
along with ``Lena,'' ``Barbara'' and ``Peppers'' used in the following
experiments), and then measuring the final codelength and computation
time. For the backward case, the initial dictionary is the global dictionary
learned in the previous experiment.  As for the forward method, we also
include a faster ``partial update'' variant which performs a few ($10$)
iterations of Algorithm~\ref{alg:mdl-dictionary-learning} after adding a new
atom, instead of allowing it to converge. The backward method produced a
dictionary of size $\natoms=170$, yielding a compression level of $5.13$bpp
at a computational cost of $3900$s. For the convergent forward method, a
dictionary with $\natoms=34$, yielding $5.19$bpp and requiring
$800$s. Finally, the forward method resulted in a dictionary of size
$\natoms=20$, $5.22$bpp, and required $150$s. In all cases, the running
times were measured for a parallelized C++ implementation running on an
Athlon Phenom II X6 at 2.6GHz).  In summary, all three methods reach
similar, significant, compression levels. Slightly better results are
obtained with the backward method, at the cost of a significant increase in
computational time. On the other hand, the partial forward variant is
significantly faster than the other two, yielding similar codelengths.
 


\subsection{Denoising of natural images}
\label{sec:results:denoising}

The task in this case is to estimate a clean image from an observed noisy
version whose pixels are corrupted by AWGN of known variance
$\sigma^2_\err$.  Here $\datam$ contains all (overlapping) $8{\times}8$
patches from the noisy image. The denoising algorithm proceeds in two
stages. In the first one, a dictionary $\dictm$ is learned from the noisy
image patches $\datam$. We use the backward selection algorithm since it
allows us to use the global dictionary as the starting point, a common
practice in this type of problems,~\cite{aharon06,mairal08a}. Secondly, the
clean patches are estimated as sparse combinations of atoms from
$\dictm$. In our case, the second stage admits two variants. The first one
is a rate-distortion (RD) procedure akin to the traditional method used for
example in~\cite{aharon06}, where each clean sample $\opt\datav_j$ is
estimated using a distortion-constrained formulation. In our case, we
minimize the codelength (or ``rate'') of describing $\datav_j$ up to a
prescribed distortion proportional to the noise level, $ \opt \datav_j =
\dictm\opt\coefv_j, \opt\coefv_j = \arg\min_\auxv L(\auxv) \st
\norm{\datav_j-\dictm\auxv}_2 \leq C\sigma^2_\noiserv.  $ Here we use
$C=1.0$. The second variant, coined ``post-thresholding'' (PT) is more
consistent with the learning phase, and is truly parameter-free, since the
estimation derives from the same codelength minimization procedure used for
learning the dictionary $\dictm$. In this case we obtain an initial estimate
$ \tilde \datav_j = \dictm\tilde\coefv_j,\; \tilde\coefv_j = \arg\min_\auxv
L(\auxv) + L(\datav_j|\auxv).  $ However, according to the model developed
in Section~\ref{sec:encoding:error}, the encoding residual
$\tilde\errv=\datav_j-\tilde\datav_j$ may contain a significant portion of
clean data due to modeling errors. We can then think of $\tilde\errv$ as
clean data corrupted by noise of variance $\sigma_\err^2$. To extract the
clean portion, we solve another codelength-minimization sub-problem, this
time with a Gaussian prior for the error, and a Laplacian prior for the
clean part, $ \bar \errv_j = \arg\min_\aux
\frac{1}{\sigma_\err^2}\norm{\tilde\errv_j-\auxv}_2 + \frac{1}{\hat
  \theta_\err } \norm{\auxv}_1, $ where $\hat\theta_\errv = \sqrt{0.5 \max
  \{0,\fun{var}(\tilde\errv_j)-\sigma_\err^2 \} }$, following
Section~\ref{sec:encoding:collaborative}.  We then compute the final
estimate as $\hat\datav_j = \tilde\datav_j + \bar\errv_j$.  In either
variant, the model used for $L(\coefv)$ includes the Markovian dependency
between the occurrence of each atom in a patch and its previously-encoded
neighbors, as described in Section~\ref{sec:encoding:collaborative}.

\begin{figure*}\figspec
\begin{center}
\begin{minipage}{0.293\textwidth}
\resizebox{\textwidth}{!}{%
\begin{tabular}[b]{|l|cccc|}\hline
$\sigma_\err=10$ &  PT & RD & \cite{roos09} & \cite{aharon06}  \\\hline
lena	& 34.9	& 35.2	& 32.4 & 35.5 \\ 
barbara	& 33.0	& 33.8	& 29.4 & 34.4 \\ 
boat	& 33.1	& 33.2	& 30.5 & 33.6 \\ 
peppers	& 34.1	& 34.4	& 32.2 & 34.3 \\\hline  
$\sigma_\err=20$ & PT & RD & \cite{roos09} & \cite{aharon06}  \\\hline
lena    & 32.0	& 32.2	& 29.4 & 32.4 \\
barbara & 29.7	& 30.6	& 25.7 & 30.8 \\
boat    & 29.5	& 30.3	& 27.5 & 30.3 \\
peppers & 31.7	& 31.6	& 29.4 & 30.8 \\ \hline
\end{tabular}%
}
\end{minipage} %
\begin{minipage}{0.697\textwidth}
\includegraphics[width=0.245\textwidth]{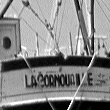}\hspace{1pt}%
\includegraphics[width=0.245\textwidth]{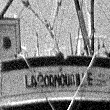}\hspace{1pt}%
\includegraphics[width=0.245\textwidth]{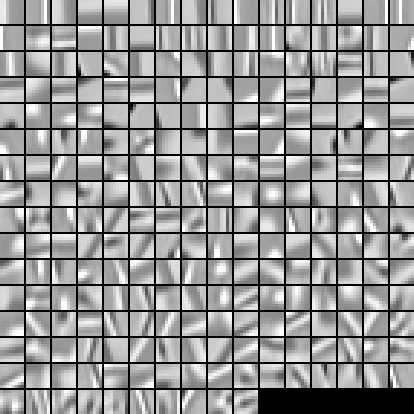}\hspace{1pt}%
\includegraphics[width=0.245\textwidth]{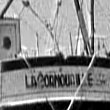}\\[1pt]
\includegraphics[width=0.245\textwidth]{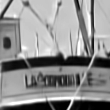}\hspace{1pt}%
\includegraphics[width=0.245\textwidth]{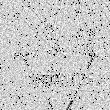}\hspace{1pt}%
\includegraphics[width=0.245\textwidth]{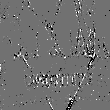}\hspace{1pt}%
\includegraphics[width=0.245\textwidth]{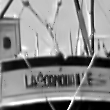}\hspace{1pt}
\end{minipage}
\vspace{-0ex}\caption{\label{fig:denoising}Denoising results. 
Left table: denoising performance, in PSNR, of K-SVD ~\cite{aharon06}, MDL denoising~\cite{roos09}, and the
   Post-Thresholding (PT) and Rate-Distortion (RD) denoising variants.
Images, top row: clean ``Boats'', noisy version, learned dictionary for this
  image (final $\natoms=248$), image recovered using RD.  Images, bottom row: image
  reconstructed from the initial estimation $\tilde\datav_j$ obtained in the
  PT method, its residual, portion of residual that was added back, final
  PT estimation.}
\end{center}
\end{figure*}

Denoising performance is summarized in Figure~\ref{fig:denoising}, along
with a detail of the result for $\sigma_\err=10$ for the ``Boats'' image in
Figure~\ref{fig:denoising}. In all cases, there is a $1$ to $5$ dB
improvement over the best MDL-based results in \cite{roos09}, thus showing
the relevance of overcoming the limitations in previous MDL applications to
sparse coding. Both the RD and PT methods yield results which are comparable
to those of \cite{aharon06}, which depend significantly on several carefully tuned
parameters.\footnote{To the best of our knowledge, these results, as well as
  those in \cite{aharon06}, are among the best that can be obtained for
  gray-scale images without using multi-scale and/or spatial aggregation of
  patches as in \cite{dabov07b,mairal09iccv}.} While the RD variant performs
better than PT in terms of PSNR, PT is faster and tends to produce less
artifacts than RD, thus resulting in more visually pleasant images than
RD. This, which can be clearly seen in Figure~\ref{fig:denoising}, occurs in
all other cases as well. Including the Markov dependency in
$L(\coefv)$ produced an average improvement of up to $0.2$dB.

\subsection{Texture mosaic segmentation}
\label{sec:results:texture}

Here we are given $\nclasses$ images with sample textures, and a target
mosaic of textures,\footnote{Taken from
  \small\url{http://www.ux.uis.no/~tranden/}.} and the task is to assign
each pixel in the mosaic to one of the textures. Again, all images are
decomposed into overlapping patches. This time a different dictionary is
learned for each texture using patches from corresponding training
images. In order to capture the texture patterns, a patch width $w=16$ was
used. Then, each patch in the mosaic is encoded using all available
dictionaries, and its center pixel is assigned to the class which produced
the shortest description length for that patch.

\begin{figure*}\figspec
\begin{center}
\includegraphics[width=0.975\textwidth]{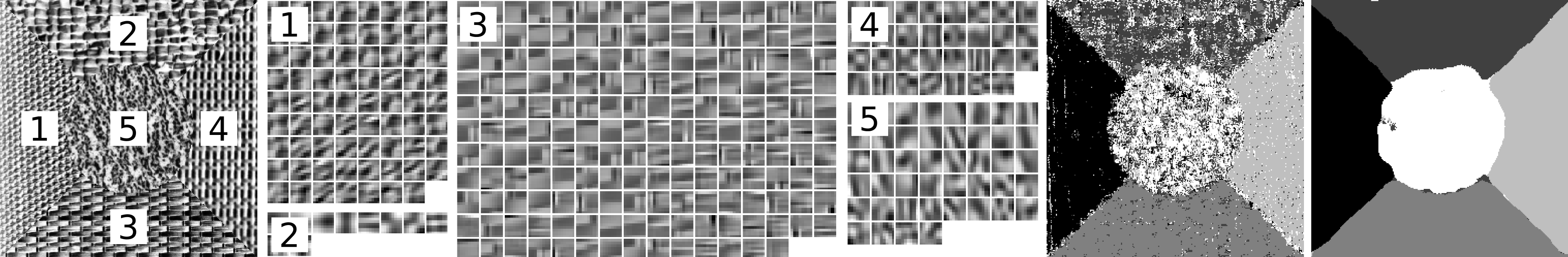}
\vspace{-0ex}\caption{\label{fig:texture-segmentation} Left to right: Texture mosaic,
  dictionaries learned for each class (note the automatically learned
  different sizes), patch-wise codelength-based classification map --each
  shade of gray corresponds to a texture class -- ($77.0\%$ success rate),
  classification map obtained by averaging the codelength over a
  neighborhood of patches ($95.4\%$ success rate). }
\end{center}
\end{figure*}

This seemingly natural procedure results in a success rate of $77\%$, which
is consistent with the second picture of
Figure~\ref{fig:texture-segmentation}. The problem is that this procedure is
inconsistent with the learning formulation, because each dictionary is
adapted to minimize the \emph{average} codelength of describing each patch
in the respective texture. Therefore, good results can only be expected if
the decision is made for groups of patches simultaneously, that is, by
considering the cumulative codelength of a set of patches. We implement this
by deciding on each patch on the basis of comparing the average codelength
obtained with each dictionary for encoding that patch and all patches in a
circular neighborhood with a radius of $20$ pixels. The success rate in this
case is $95.3\%$, which is comparable to the state-of-the art for this type
of problems (see for example~\cite{mairal08c}, which learns sparse models
for explicitly maximizing the success rate).  The Markovian model improved
our results by $1\%$.

\subsection{Low-rank matrix approximation}
\label{sec:results:low-rank}

The low-rank matrix approximation family of problems (see~\cite{candes11acm}
for a review) can be seen as an extension to the problem of sparse coding
where sparsity is substituted by matrix rank. Concretely, the task is to
recover a matrix $\coefm \in\reals^{\ndims{\times}\nsamples}$ from an
incomplete and/or corrupted observation $\datam$, under the assumption that
the rank of $\coefm$, $\rank(\coefm)$, is small. As with sparse coding,
$\rank(\coefm)$ is relaxed using the \cost{1} equivalent for matrix rank,
which is the nuclear norm, $\norm{\coefm}_* \defeq \sum_i \sigma_i(\coefm)$,
where $\sigma_i(\coefm)$ is the $i$-th singular value of $\coefm$.  It has
been shown in \cite{candes11acm} that, under certain assumptions on
$\rank(\coefm)$, the following estimation function is able to recover
$\coefm$ from a noisy observation $\datam$, and with a significant fraction
of its coefficients arbitrarily corrupted,
\begin{equation}
\hat\coefm = \arg\min_{\mat{W}} \norm{\mat{W}}_* + \lambda \norm{\datam-\mat{W}}_1,\quad \lambda=1/\sqrt{\max \{\ndims,\nsamples\}}.
\label{eq:rpca}
\end{equation}

A common proof of concept is to use this framework for robust background
estimation in camera surveillance video sequences~\cite{wright09nips}, and
we apply our proposed framework for the same application.

To perform our MDL-based model selection within this formulation, we solve
\refeq{eq:rpca} for increasing values of $\lambda$, obtaining a low-rank
approximation to $\coefm$,
$(\coefm(\lambda),\errm(\lambda)=\datam-\coefm(\lambda))$, which we encode
using the universal models described in
Section~\ref{sec:encoding-scheme}. We modified the algorithm described in
\cite{lin09arxiv} to allow for warm restarts, using the solution for the
previous $\lambda$ as a starting point for the next $\lambda$ for faster convergence.

Consistently with the $\cost{1}$ fitting term of \refeq{eq:rpca}, we encode
the non-zero values of $\errm(\lambda)$ as a Laplacian sequence of unknown
parameter. To exploit the potential sparsity in $\errm(\lambda)$, the
locations of the non-zero values are encoded, as
in~\ref{sec:encoding:coefficients}, using an enumerative two-parts code for
Bernoulli sequences of unknown parameter.  To exploit low-rank in the
encoding, the matrix $\coefm(\lambda)$ is encoded via its reduced SVD
decomposition $\coefm(\lambda) = \mat{U}(\lambda) \Sigma(\lambda)
\mat{V}(\lambda)\transp$.  For $\rank(\coefm(\lambda))=r$, we have that
$\mat{U}(\lambda) \in \reals^{\ndims{\times}r}$ are the left-eigenvectors, $\Sigma
\in \reals^{r{\times}r}$ is the diagonal matrix whose diagonal are the
non-zero singular values of $\coefm(\lambda)$, and $\mat{V}(\lambda) \in
\reals^{r{\times}\nsamples}$ are the right-eigenvectors of
$\coefm(\lambda)$.  Each column of $\mat{U}$ is encoded (in this video
example) as a smooth image via a causal bilinear predictor identical to the
one used for predictive coding of $\dictm$ in~\ref{sec:encoding:dictionary},
using a Laplacian model for the prediction residuals. Each column of
$\mat{V}$ is encoded as a smooth one-dimensional sequence, using a zero
order predictor (the predicted value for the next coefficient is the
previous coefficient value), with a Laplacian prior on the prediction
residuals.  Finally, the values of $\mat{\Sigma}$, which can be arbitrary,
are quantized and encoded using the universal code for
integers~\cite{rissanen83}.

The encoding method is very simple, with all unknown parameters encoded
using a two-parts code, and codelenghts for the discretized Laplacian
pre-computed in look-up tables. Quantization for this case is as follows:
the codelength associated with the $r$ non-zero singular values is
negligible, and we minimize unwanted distortion encoding them with high
precision ($1e-16$). As for the columns of $\mat{U}$ and $\mat{V}$, they all
have unit norm, so that the average magnitude of their elements are close to
$\sqrt{1/\ndims}$ and $\sqrt{1/\nsamples}$ respectively. Based on this, our
algorithm encodes the data with $\delta_u=Q/\sqrt{\ndims}$ as the
precision for encoding $\mat{U}$, and $\delta_v=Q/\sqrt{\ndims}$ for
$\mat{V}$, for several values of $Q$ in $(0,1)$, keeping the one producing
 the smallest codelength.
The MDL-based estimation algorithm then chooses the model for which
the codelength $L(\datam;\lambda) = L(\mat{U}(\lambda)) + L(\Sigma(\lambda))
+ L(\mat{V}(\lambda))$ is minimized.

\begin{figure*}\figspec
\begin{center}
\subfloat[\label{fig:low-rank-a}Results for ``Lobby'' sequence, featuring a
  room with lights that are switched off and on. The rank of the
  approximation for this case is $\rank=10$.  The moment where the lights
  are turned off is clearly seen here as the ``square pulse'' in the middle
  of the first two right-eigenvectors (bottom-right).  Also note how
  $\vec{u}_2$ (top-right) compensates for changes in
  shadows.]{\includegraphics[width=0.485\textwidth]{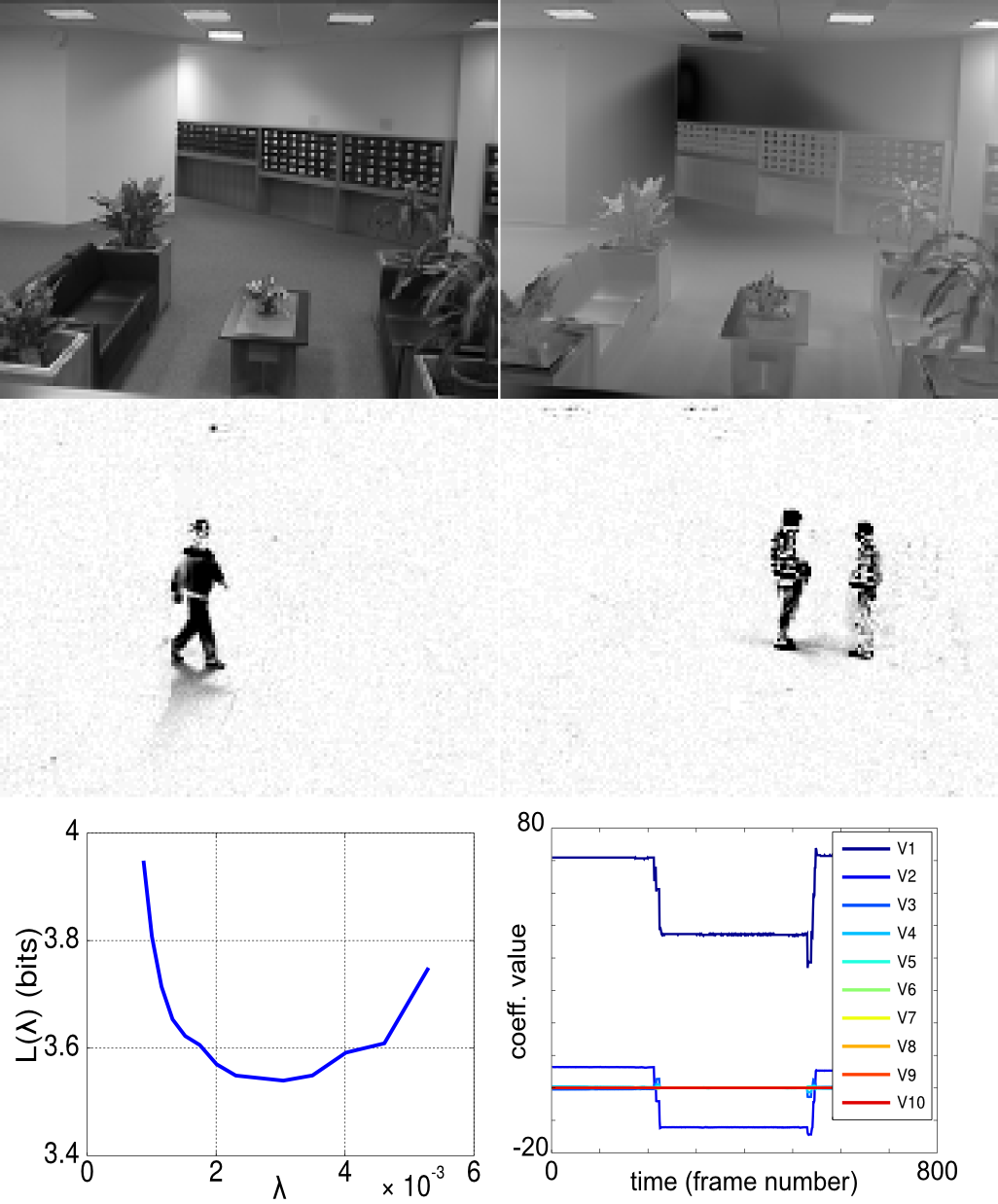}}\hspace{2ex}%
\subfloat[\label{fig:low-rank-b}Results for ``ShoppingMall'', a fixed camera
  looking at a crowded hall. In this case, the rank of the approximation
  decomposition is $\rank=7$. Here, the first left-eigenvector models the
  background, whereas the rest tend to capture people that stood still for a
  while. Here we see the ``phantom'' of two such persons in the second
  left-eigenvector
  (top-right).]{\includegraphics[width=0.485\textwidth]{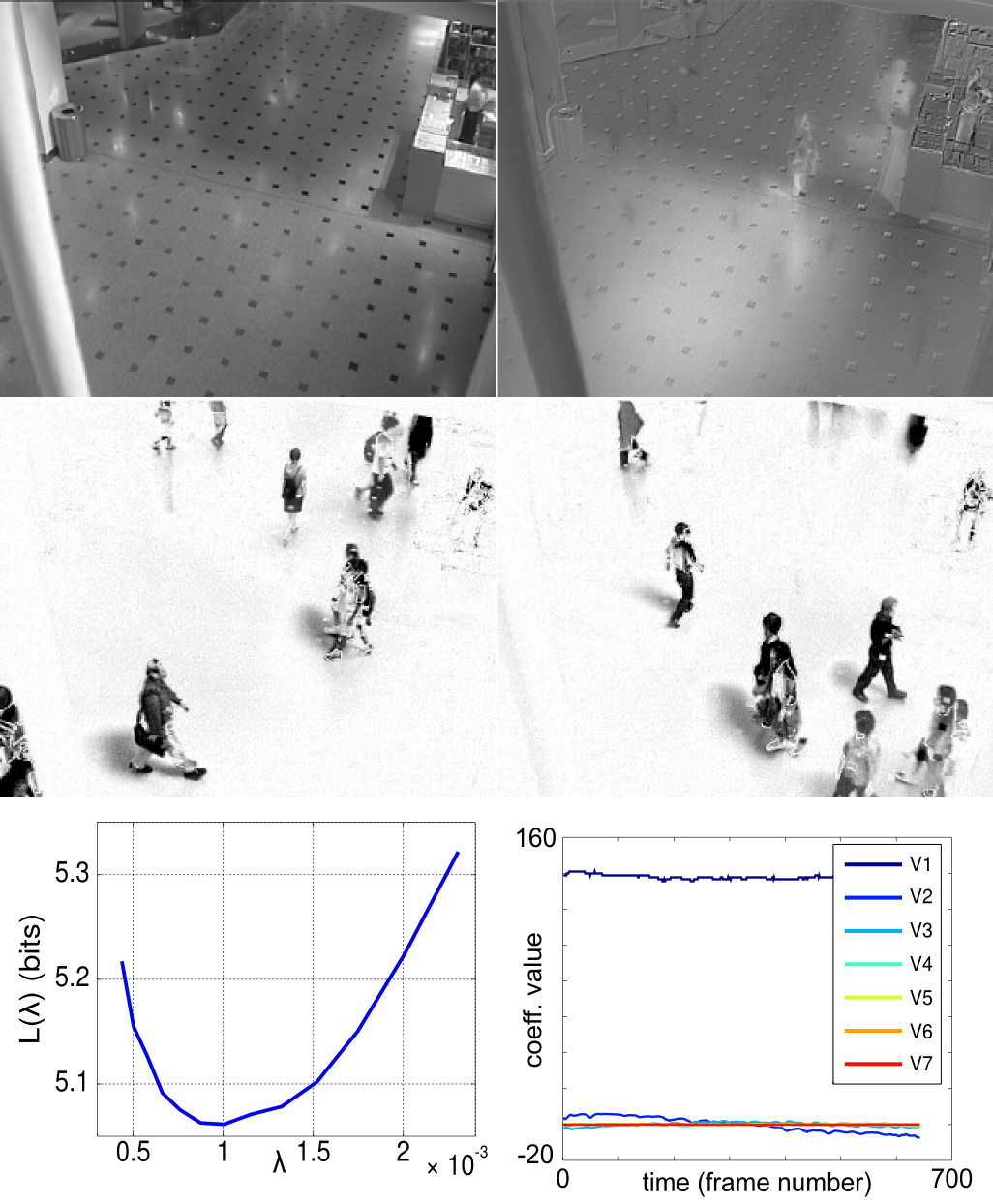}}%
\vspace{-1ex}\caption{\label{fig:low-rank}Low-rank approximation results.  Both figures
  show the first two left-eigenvectors as 2D images at the top, two sample
  frames from the approximation error sequences in the middle, which should
  contain the people that were removed from the videos, and the curve
  $L(\lambda)$ and the right-eigenvalues, scaled by $\Sigma$ (representing
  the ``activity'' of each left-eigenvector along time), at the bottom. }
\end{center}
\end{figure*}

As in \cite{wright09nips}, here we show results for two sequences taken from
\cite{li04tip}: ``Lobby'' (Figure~\ref{fig:low-rank}(a)), and
``ShoppingMall'' (Figure~\ref{fig:low-rank}(b)).  Full videos can be viewed at
{\small\url{http://www.tc.umn.edu/~nacho/lowrank/}}.

In both cases, the recovered backgrounds are very
accurate. In particular, for the Lobby sequence, the selected model captures
just the eigenvectors needed to recover the background along with its
lighting changes, including corrections for local shadows, leaving out only
the people passing by.

\section{Concluding remarks}
\label{sec:conclusion}

We have presented an MDL-based sparse modeling framework, which
automatically adapts to the inherent complexity of the data at hand using
codelength as a metric. 

The framework features a sparse coding algorithm and
automatic tuning of the sparsity level on a per-sample basis, including a
sequential collaborative variant which adapts the model parameters as it
processes new samples, and two dictionary learning variants which learn the
size of the dictionaries from the data. In all cases, the
information-theoretic formulation led to robust coding and learning
formulations, including novel robust metrics for the fitting term (LG and
MOEG), and robust $\cost{1}$-based dictionary regularization term.  This
formulation also allowed us to easily incorporate more prior information
into the coding/learning process, such as Markovian spatial dependencies, by
means of simple modifications to the probability models used.

As a result, the framework can be applied out-of-the-box to very different
applications, from image denoising to low-rank matrix approximation,
obtaining competitive results in all the cases presented, with minimal
interaction from the user.

\balance
\bibliography{sparse}
\bibliographystyle{IEEEbib}

\end{document}